\documentclass[12pt]{article}
\usepackage[
    a5paper,
    left=1cm,
    right=1cm,
    top=1.5cm,
    bottom=1cm,
    nomarginpar, nohead, includefoot
]{geometry}
\usepackage{biblatex, comment, mathtools}
\usepackage{amssymb, amsmath, amsthm}
\usepackage{stmaryrd, extarrows, mathrsfs}
\usepackage{mdframed, float}
\usepackage[bw]{agda}
\usepackage{tikz, quiver, mathpartir}
\usepackage{xcolor, xurl, hyperref, cleveref}

\floatstyle{boxed}
\restylefloat{figure}

\hypersetup{
    colorlinks = true,
    linkcolor = {teal},
    anchorcolor = {teal},
    citecolor = {violet},
    urlcolor = {cyan!70!black}
}

\newtheoremstyle{idea}
  {-\topsep}%                 <space above>
  {}%                         <space below>
  {\normalfont}%              <body font>
  {}%                         <indent amount>
  {\bfseries}%                <theorem head font>
  {.}%                        <punctuation after theorem head>
  {.5em}%                     <space after theorem head>
  {}%                         <theorem head spec>
\theoremstyle{idea}
\newmdtheoremenv[linecolor=cyan,
  backgroundcolor=cyan!10]
  {idea}{Idea}

\theoremstyle{definition}
\newtheorem{theorem}{Theorem}[section]
\newtheorem{lemma}[theorem]{Lemma}
\newtheorem{definition}[theorem]{Definition}
\newtheorem*{remark}{Remark}
\AtBeginEnvironment{remark}{%
  \pushQED{\qed}%
}
\AtEndEnvironment{remark}{\popQED\endremark}

\DeclareFontFamily{U}{min}{}
\DeclareFontShape{U}{min}{m}{n}{<-> udmj30}{}
\DeclareMathOperator{\yo}{\!\text{\usefont{U}{min}{m}{n}\symbol{'210}}\!}

\newcommand{\explain}[1]{\text{\emph{\color{gray}{#1}}}}

\newcommand{\C}[1]{\textsf{#1}}
\newcommand{\op}{^{\textsf{op}}}
\newcommand{\CommaCat}{\mathbin{\downarrow}}
\newcommand{\atom}[1]{\texttt{'#1}}

\newcommand{\Of}{\mathord{:}}
\newcommand{\of}{\mathrel{\Of}}
\newcommand{\isType}{\;\textnormal{type}}

\definecolor{varcolor}{RGB}{140, 140, 140}
\newcommand{\isVar}{\mathrel{\color{varcolor}\Of_{\textnormal{var}}}}
\definecolor{nfcolor}{RGB}{30, 150, 200}
\newcommand{\isNf}{\mathrel{\color{nfcolor}\Of_\textnormal{nf}}}
\newcommand{\isNfType}{\;{\color{nfcolor}\textnormal{type}_\textnormal{nf}}}
\definecolor{necolor}{RGB}{190, 40, 120}
\newcommand{\isNe}{\mathrel{\color{necolor}\Of_\textnormal{ne}}}
\newcommand{\isNeType}{\;{\color{necolor}\textnormal{type}_\textnormal{ne}}}

\newcommand{\Unit}{\mathbf{1}}
\newcommand{\Ans}{\textsf{Ans}}
\newcommand{\yes}{\textsf{yes}}
\newcommand{\no}{\textsf{no}}
\newcommand{\Bool}{\textsf{Bool}}
\newcommand{\true}{\textsf{true}}
\newcommand{\false}{\textsf{false}}
\newcommand{\caseof}{\mathop{\textsf{case}}}

\newcommand{\Univ}{\textsf{U}}
\newcommand{\Ucode}[1]{\texttt{#1}}
\DeclareMathOperator{\UEl}{\textsf{El}}

\newcommand{\something}{\mathord{\mkern1mu{\color{gray}%
  \rule[-0.5pt]{0.35em}{0.3pt}%
}\mkern1mu}}
\newcommand{\bind}{\mathpunct{.}}

\DeclarePairedDelimiter\Sem{\llbracket}{\rrbracket}
\DeclarePairedDelimiter\Ren{\llparenthesis}{\rrparenthesis}
\DeclareMathOperator{\reify}{\Downarrow}
\DeclareMathOperator{\reflect}{\Uparrow}
\DeclareMathOperator{\semantics}{\textsc{eval}}
\DeclareMathOperator{\normalize}{\textsc{nbe}}

\DeclareMathOperator{\GSect}{\boldsymbol{\Gamma}}
\DeclareMathOperator{\Nerve}{\mathbf{N}}

\newcommand{\Type}{\textnormal{Tp}}
\newcommand{\Typenf}{\textnormal{Tp}_{\color{nfcolor}\textnormal{nf}}}

\newcommand{\Term}{\textnormal{Tm}}
\newcommand{\Termvar}{\textnormal{Tm}_{\color{varcolor}\textnormal{var}}}
\newcommand{\El}{\textnormal{El}}
\newcommand{\Elnf}{\textnormal{El}_{\color{nfcolor}\textnormal{nf}}}
\newcommand{\Elne}{\textnormal{El}_{\color{necolor}\textnormal{ne}}}
\newcommand{\Elvar}{\textnormal{El}_{\color{varcolor}\textnormal{var}}}
\newcommand{\synct}[1]{\textbf{\textsf{#1}}}
\newcommand{\nfct}[1]{\textbf{\textsf{\color{nfcolor}{#1}}}}
\newcommand{\nect}[1]{\textbf{\textsf{\color{necolor}{#1}}}}
\newcommand{\typeof}{\textnormal{typeof}}
\newcommand{\typeofvar}{\textnormal{typeof}_{\color{varcolor}\textnormal{var}}}

\newcommand{\comp}{\mathbin{\circ}}

\newcommand{\syn}{\circ}
\newcommand{\sem}{\bullet}
\newcommand{\Psyn}{{\mathord{\textnormal{\S}}}}
\newcommand{\Pleft}{{\mathord{\textnormal{\S}}_L}}
\newcommand{\Pright}{{\mathord{\textnormal{\S}}_R}}
\newcommand{\Casing}[2]{\left[ \Pleft \hookrightarrow {#1} ; \Pright \hookrightarrow {#2}\right]}

\newcommand{\Msyn}{%
\mathopen{\mkern1mu%
\text{\tikz[baseline={($(current bounding box.south) + (0, 0.3pt)$)}]{%
  \draw[line width=0.56pt] circle(0.79ex);%
}}\mkern1mu}}
\newcommand{\Msem}{%
\mathopen{\mkern1mu%
\text{\tikz[baseline={($(current bounding box.south) + (0, 0.3pt)$)}]{%
  \filldraw[line width=0.56pt] circle(0.79ex);%
}}\mkern1mu}}

\newenvironment{internal}
{\vskip -3pt\mdframed[hidealllines=true,
backgroundcolor=gray!10,
innertopmargin=3pt,
innerbottommargin=3pt,
innerleftmargin=0pt,
innerrightmargin=0pt,
splitbottomskip=3pt,
splittopskip=14pt]
\AgdaNoSpaceAroundCode
\code}
{\endcode\endmdframed\vskip -3pt}

\renewcommand{\AgdaFontStyle}[1]{\textnormal{#1}}
\newcommand{\kw}[1]{\textbf{#1}}
\newcommand{\ct}[1]{\textsf{#1}}
\newcommand{\pr}[1]{\textsf{#1}}
\newcommand{\vb}[1]{\textit{#1\/}}

\newcommand{\iType}{\mathscr{U}}
\newcommand{\ext}[3][\Psyn]{\left\{%
  {#2}\,\middle|\, {#1} \hookrightarrow {#3}%
\right\}}
\newcommand{\extt}[3]{\left\{%
  {#1}\,\middle|\, \Pleft \hookrightarrow {#2} ; \Pright \hookrightarrow {#3}%
\right\}}
\newcommand{\Refine}{\mathop{\vphantom{\sum}\mathchoice
  {\vcenter{\hbox{\huge G}}}
  {\vcenter{\hbox{\Large G}}}
  {\mathrm{G}}
  {\mathrm{G}}}}
\DeclareMathOperator{\Comp}{Comp}
\newcommand{\hole}{\colorbox{yellow!90!green}{\textsf{?}}}

\newenvironment{logical}
{\vskip -3pt\mdframed[hidealllines=true,
backgroundcolor=green!10,
innertopmargin=3pt,
innerbottommargin=3pt,
innerleftmargin=0pt,
innerrightmargin=0pt,
splitbottomskip=3pt,
splittopskip=14pt]
\AgdaNoSpaceAroundCode
\code}
{\endcode\endmdframed\vskip -3pt}

\newcommand{\Jdg}{\mathord{\square}}

\begin{document}
% \sloppy
\title{Synthetic Tait Computability\\The Hard Way}
\author{Huang Xu}
\date{}
\maketitle
\begin{abstract}
We walk through a few proofs of canonicity and normalization,
each one with more aspects dissected and re-expressed in category theory,
so that readers can compare the difference across proofs.
During this process we isolate the different ideas that make up the proofs.
Finally we arrive at \emph{synthetic Tait computability} as proposed by J.\@ Sterling.
We also give a synthetic proof for parametricity of system F.
\end{abstract}

\setcounter{section}{-1}  % Start at zero
\section{Introduction}
Tait computability, also known as logical relations,
plays an important role in establishing metatheorems about type theories.
However, different treatments of this technique
take place in different type systems,
and are often mixed with introductions of other ideas.
This may be overwhelming for learners.
These notes are the product of the author trying to make sense of all these
and to come up with a coherent web structuring the ideas.

In these notes, we spell out several proofs using Tait computability,
introducing each idea separately.
We also prove some theorems repeatedly,
each time using more of these ideas.
We assume familiarity with the lambda calculus.
In particular, we will only present the naïve proofs
to fix notation and to compare with later proofs,
not offering any explanation.
For an in-depth introduction, see \textcite{skorstengaard:2019:logrel}.

We require some experience with type-theoretic proof assistants such as Agda.
Some familiarity with categories is also needed,
but we will try to introduce any needed categorical machinery.
We will throw some algebraic geometry words here and there
but they're included just for fun.

\section{Simple Types, Naïvely}

\subsection{Warming up}

Let's consider a calculus with unit, product and function types.
We assume de Bruijn variables, but use named variables in our notation.
We also add a type with two elements, for demonstration.
\begin{mathpar}
\inferrule{ }{\Ans\isType} \and
\inferrule{ }{\Gamma \vdash \yes \of \Ans} \and
\inferrule{ }{\Gamma \vdash \no \of \Ans}
\end{mathpar}
We don't have eliminators for this type.
Hence this is not the boolean type.
However, these two elements are still distinct, judgementally.
\begin{theorem}[Equational consistency]
We cannot derive \[\Gamma \vdash \yes = \no \of \Ans.\]
\end{theorem}
\begin{proof}
We assign a set \(\Sem{A}\) to every type,
a set \(\Sem{\Gamma}\) to every context,
and a function \(\Sem{a} \of \Sem{\Gamma} \to \Sem{A}\)
to every term \(\Gamma \vdash a \of A\).
We define the assignment inductively,
keeping the invariant that \(\Gamma \vdash a = b \of A\)
implies \(\Sem{a} = \Sem{b}\).

We assign \(\Sem{\Unit} = \{\atom{nil}\}\)%
\footnote{Expressions such as \(\atom{nil}\) are \emph{atoms}.
You may take them as anything, as long as atoms with different names are distinct.
For example, you may define \(\atom{nil}\) to be the natural number \(57\),
and \(\atom{mil} = 58\), etc.}
an arbitrary singleton set.
Product types \(\Sem{A \times B} = \Sem{A} \times \Sem{B}\)
are interpreted as the Cartesian product.
Function types are interpreted as function sets
\(\Sem{A \to B} = \Sem{A} \to \Sem{B}\).
Finally we interpret \(\Sem{\Ans} = \{\atom{t}, \atom{f}\}\)
as a two-element set.

Contexts are simply lists of types.
We assign them the Cartesian product of the interpretation of types.
In other words, the elements of \(\Sem{\Gamma}\) are lists
\(\vec u\) whose \(k\)-th element belongs to \(\Sem{A_k}\),
where \(A_k\) is the \(k\)-th type in \(\Gamma\).
This induces corresponding projection functions
\(\Sem{\Gamma} \to \Sem{A_k}\),
which we take as the interpretation of variables.
We can recursively interpret the terms.
The unique element of the unit type is interpreted as
the unique function \(\Sem{\Gamma} \to \{\atom{nil}\}\).
\begin{align*}
\Sem{(a,b)}(\vec u) &= (\Sem{a}(\vec u), \Sem{b}(\vec u)) \\
\Sem{\pi_1(p)} &= \pi_1 \comp \Sem{p} \\
\Sem{\pi_2(p)} &= \pi_2 \comp \Sem{p} \\
\Sem{\lambda x\bind t}(\vec u) &= u' \mapsto \Sem{t}(\vec u, u') \\
\Sem{t(s)}(\vec u) &= \Sem{t}(\vec u)(\Sem{s}(\vec u)) \\
\Sem{\yes}(\vec u) &= \atom{t} \\
\Sem{\no}(\vec u) &= \atom{f}
\end{align*}
The \(\pi_1, \pi_2\) on the right hand side are set-theoretic functions,
not related to our syntax.
We can inductively verify that all the equations are indeed satisfied.
So we have \(\Gamma \vdash a = b \of A \implies \Sem{a} = \Sem{b}\).
Taking the contrapositive, if \(\Sem{a} \ne \Sem{b}\),
then the corresponding judgemental equality cannot be derived.
This finishes the proof.%
\footnote{There is a slight concern that
\(\Sem{\Gamma}\) might be empty, and functions with empty domain are all equal.
This possibility can be quickly disposed of by induction.}
\end{proof}
\begin{remark}
The theorem is different from \emph{logical consistency},
which takes the empty type and states that we cannot construct
any element of this type in the empty context.
This is an easy scholium of our canonicity result later,
and we trust the reader to give a proof.
\end{remark}

\subsection{Canonicity}

How do we confirm that we haven't forgotten any judgemental equalities?
One way to check that is to prove \emph{canonicity}.
We can state canonicity for our type theory as
saying every closed term of \(\Ans\) is equal to either \(\yes\) or \(\no\).
Additionally, we can require
an effective algorithm that decides which case it is.
This then gives a \emph{computational meaning} to our type theory.
In other words, our type theory can be regarded as a type system for
\emph{programs} that compute to a \emph{value}.

Before we dive into the proof let's reflect on
why our previous construction doesn't fit the bill.
In the set-theoretic interpretation,
we interpreted \(\Ans\) as a two-element set,
so the interpretation of a term \(\vdash t \of \Ans\)
lands in the two-element function set \(\Sem{()} \to \Sem{\Ans}\).
We proved that judgementally equal elements
have set-theoretically equal interpretations,
but not the converse.
Therefore it may be the case that
there are other exotic closed terms of \(\Ans\)
that happen to have the same interpretation with \(\yes\) or \(\no\),
but nevertheless are judgementally different.
Hence we need to modify the proof to track more invariants.
The data structures used for such tracking are called \emph{computability structures}.

\begin{theorem}[Canonicity]
For every term \(\vdash t \of \Ans\),
either \(\vdash t = \yes \of \Ans\) or \(\vdash t = \no \of \Ans\) is derivable.
\end{theorem}
\begin{proof}
For each closed term \(\vdash t \of A\) considered up to judgemental equality,
we define a set of computability structures over it
via inducion on the type.
In particular, we arrange so that
there exists a computability structure over \(\vdash t \of \Ans\)
iff it is equal to \(\yes\) or \(\no\).
The computability structure for other terms are defined
so that the induction goes through.
Finally we inductively choose a computability structure for each term.

For the product type,
we define a computability structure for \(\vdash p \of A \times B\)
as a pair whose first element is a computability structure for \(\pi_1(p)\),
and the second element is a computability structure for \(\pi_2(p)\).
For the unit type,
we define the set of computability structures for any element
as a singleton set \(\{\atom{nil}\}\).
For the function type,
we define a computability structure for \(\vdash f \of A \to B\) as a function.
Its domain consists of pairs \((a, u)\) where \(\vdash a \of A\)
and \(u\) is a computability structure for \(a\).
It maps each \((a, u)\) to a computability structure for \(\vdash f(a) \of B\).
Finally, we define a computability structure for \(\vdash t \of \Ans\) as
\[\{\atom{t} \mid t = \yes\} \cup \{\atom{f} \mid t = \no\}.\]

Although the result does not mention open terms,
we still need to deal with them
since \(\lambda\) introduces variables into the context.
We define computability \emph{morphisms} for \(\Gamma \vdash t \of B\)
as maps whose domain is \((\vec a, \vec u)\) where
\(\vec a\) is a list of closed terms with type given by \(\Gamma\),
and \(\vec u\) is a list of equal length
containing corresponding computability structures.
These maps are required to map \((\vec a, \vec u)\)
to computability structures for \(t[\vec x/\vec a]\).
Observe that for a closed term \(t\), its computability morphisms
are in bijective correspondence with its computability structures.

We perform induction on the terms
to select a computability morphism \(\Sem{t}\) for each term,
and hence select computability structures for closed terms.
Recall that terms are considered up to judgemental equality.
So we additionally need to prove that whenever \(\Gamma \vdash t = s \of A\),
we have \(\Sem{t} = \Sem{s}\).
Interpretation for variables are formed similarly by projections.
\begin{align*}
\Sem{\yes}(\vec a, \vec u) &= \atom{t}\\
\Sem{\no}(\vec a, \vec u) &= \atom{f}\\
\Sem{()}(\vec a, \vec u) &= \atom{nil}\\
\Sem{(s, t)}(\vec a, \vec u) &= (\Sem{s}(\vec a, \vec u), \Sem{t}(\vec a, \vec u))\\
\Sem{\pi_1(p)} &= \pi_1 \comp \Sem{p}\\
\Sem{\pi_2(p)} &= \pi_2 \comp \Sem{p}\\
\Sem{\lambda x\bind t}(\vec a, \vec u) &= (a', u') \mapsto \Sem{t}((\vec a, a'), (\vec u, u'))\\
\Sem{t(s)}(\vec a, \vec u) &= \Sem{t}(\vec a, \vec u)(\Sem{s}(\vec a, \vec u))
\end{align*}
All the desired equation holds.
Now given \(\vdash t \of \Ans\),
\(\Sem{t}\) must be either \(\atom{t}\) or \(\atom{f}\),
and by construction this proves that it is equal to \(\yes\) or \(\no\).
\end{proof}
\begin{remark}
Traditionally, the proof of canonicity uses proof-irrelevant computability \emph{predicates}
instead of structures, or equivalently
we restrict every term to have at most one computability structure.
However computability structures are easier to generalize.
We will encounter proof-irrelevant computability in system F.
\end{remark}
\begin{remark}
The proof is completely constructive.
Therefore with a close examination of the proof
(perhaps with the aid of a proof assistant),
one can extract a concrete algorithm.
\end{remark}
It is obvious that we can and should try to figure out
exactly what data we need to provide to build the \(\Sem{-}\) thingy,
and package that into an abstract definition.
This can be done without categories,
and what it produces is called a \emph{model}.
However it's much easier with categorical tools, so we will wait until then.

\subsection{Normalization}

Finally, we wish to produce a strengthening of canonicity
that applies to all types and all contexts.
For \(\vdash t \of \Ans\), it is obvious that we want
\(\yes\) and \(\no\) in our theorem statement.
These are called \textbf{normal forms}.
The generalized theorem should state that
every term \(\Gamma \vdash t \of A\) is judgementally equivalent
to a normal form in this context and this type.

It turns out that to define normal forms,
we also need to simultaneously define
another notion called \textbf{neutral forms}.
In short, normal forms are nested constructors,
with the innermost layer being neutral;
neutral forms are eliminators nested at the eliminant,%
\footnote{The \emph{eliminant} is what the eliminator is eliminating.
For example, \(p\) is the eliminant of \(\pi_1(p)\),
and \(f\) is the eliminant of \(f(x)\).
It is called the \emph{active formula} in sequent calculus.}
with the inner most term being a variable.

More formally, we define
\(\Gamma \vdash a \isNe A\) and
\(\Gamma \vdash a \isNf A\) inductively,
in \Cref{fig:ne-nf}.
\begin{figure}
\begin{mathpar}
\inferrule{\Gamma \vdash x \isVar A}{\Gamma \vdash x \isNe A}\and
\inferrule{ }{\Gamma \vdash \yes \isNf \Ans} \and
\inferrule{ }{\Gamma \vdash \no \isNf \Ans} \and
\inferrule{\Gamma \vdash a \isNe \Ans}{\Gamma \vdash a \isNf \Ans} \and
\inferrule{ }{\Gamma \vdash () \isNf \Unit} \and
\inferrule{\Gamma \vdash a \isNf A \and \Gamma \vdash b \isNf B}
    {\Gamma \vdash (a, b) \isNf A \times B} \and
\inferrule{\Gamma \vdash p \isNe A \times B}{\Gamma \vdash \pi_1(p) \isNe A} \and
\inferrule{\Gamma \vdash p \isNe A \times B}{\Gamma \vdash \pi_2(p) \isNe B} \and
\inferrule{\Gamma, x \Of A \vdash t \isNf B}
    {\Gamma \vdash \lambda x\bind t \isNf A \to B} \and
\inferrule{\Gamma \vdash f \isNe A \to B \and \Gamma \vdash a \isNf A}
    {\Gamma \vdash f(a) \isNe B}
\end{mathpar}
\caption{Definition of neutral and normal forms}
\label{fig:ne-nf}
\end{figure}
Note that we do not allow neutral terms of product and function types to be normal.
This is to force \(\eta\) expansion.
Types like \(\Ans\) which does not have \(\eta\) rules can be seen as \emph{intensional},
and we only allow neutral forms of intensional types to be normal.

\begin{theorem}[Normalization]
There is a function mapping terms to normal forms, such that
\begin{itemize}
\item judgementally equal terms are mapped to equal normal forms;
\item terms mapped to equal normal forms are judgementally equal;
\item normal forms are fixpoints of this function.
\end{itemize}
\end{theorem}
\begin{proof}
For each term \(\Gamma \vdash t \of A\), we define
the set of computability structures over it.
For our base type \(\Ans\),
we define a computability structure for \(\Gamma \vdash t \of \Ans\)
to be the set of normal forms equal to \(t\).
For products and the unit type, the definition is the same as previous.
For the function type, there is a mysterious \emph{Kripke construction}.
We define a computability structure for \(\Gamma \vdash f \of A \to B\)
as a function.
However, its domain consists of quadruples \((\Theta, \sigma, a, u)\),
where \(\Theta \vdash \sigma \isVar \Gamma\), \(\Theta \vdash a \of A\)
and \(u\) is a computability structure for \(a\).
It maps \((\Theta, \sigma, a, u)\) to a computability structure for \(f[\sigma](a)\).

Here, we restrict \(\sigma\) so that it only substitutes variables with variables.
This is called a \textbf{renaming}.
The traditional explanation is ``it lets induction go through''.
Because $\lambda$ introduces new variables into scope,
we must allow a change of context,
but we can't allow arbitrary substitutions because
normal forms and neutral forms are not stable under substitutions.
On the contrary, they are stable under renamings.
In addition, by induction we see that
computability structures are also stable under renamings.
In other words, for a computability structure
\(u\) for \(\Gamma \vdash t \of A\)
and a renaming \(\Theta \vdash \sigma \isVar \Gamma\),
we have a corresponding computability structure
\(u[\sigma]\) for \(\Theta \vdash t[\sigma] \of A\).

We similarly define computability morphisms.
For a term \(\Gamma \vdash t \of A\) and a context \(\Delta\),
consider the set of tuples \((\vec a, \vec u)\),
where \(\vec a\) is a list of terms in the context \(\Delta\),
and the \(k\)-th term has the same type as the \(k\)-th type in \(\Gamma\);
\(\vec u\) are their corresponding computability structures.
A computability morphism over \(t\) maps these tuples to
computability structures of \(t[\vec x/ \vec a]\).
Of course, \(\vec a\) is nothing other than a substitution
\(\Delta \vdash \vec a \of \Gamma\).
But for consistency we keep our notation from our canonicity proof.

Now we do our routine induction on terms to define computability morphisms.
For clarity, we write \(\Sem{\Gamma \vdash t \of A}\)
instead of \(\Sem{t}\) to be explicit about the context and the type.
Since a computability morphism is supposed to provide a mapping
for each \(\Delta\), we write \(\Delta\) as a subscript.
Again, for variables,
\(\Sem{\Gamma \vdash x \of A}_\Delta(\vec a, \vec u)\)
just fetches the corresponding entry in \(\vec u\).
\begin{align*}
\Sem{\Gamma \vdash \yes \of \Ans}_\Delta(\vec a, \vec u) &= \yes \\
\Sem{\Gamma \vdash \no \of \Ans}_\Delta(\vec a, \vec u) &= \no \\
\Sem{\Gamma \vdash () \of \Unit}_\Delta(\vec a, \vec u) &= \atom{nil}\\
\Sem{\Gamma \vdash (s, t) \of A \times B}_\Delta(\vec a, \vec u)
    &= (\Sem{\Gamma \vdash s \of A}_\Delta(\vec a, \vec u), \Sem{\Gamma \vdash t \of B}_\Delta(\vec a, \vec u))\\
\Sem{\Gamma \vdash \pi_1(p) \of A}_\Delta(\vec a, \vec u) &= \pi_1(\Sem{\Gamma \vdash p \of A \times B}) \\
\Sem{\Gamma \vdash \pi_2(p) \of B}_\Delta(\vec a, \vec u) &= \pi_2(\Sem{\Gamma \vdash p \of A \times B}) \\
\Sem{\Gamma \vdash \lambda x\bind t \of A \to B}_\Delta(\vec a, \vec u)
    &=\\ \mathrlap{\hspace{-3cm} %{\color{gray!50} \hookrightarrow \;}
(\Theta, \sigma, a', u') \mapsto
\Sem{\Gamma, x \Of A \vdash t \of B}_{\Theta}(\vec a[\sigma], a', \vec u[\sigma], u')} \\
\Sem{\Gamma \vdash t(s) \of B}_\Delta(\vec a, \vec u)
    &=\\ \mathrlap{\hspace{-3cm} %{\color{gray!50} \hookrightarrow \;}
\Sem{\Gamma \vdash t \of A \to B}_\Delta
    (\Delta, \mathrm{id}, s[\vec x/\vec a], \Sem{\Gamma \vdash s \of A}_{\Delta}(\vec a, \vec u))}
\end{align*}
% We use \(\color{gray!60} \hookrightarrow\) to show line wraps.
Although there are loads of annotations lying around,
one can see that it still follows roughly the same structure
as in the canonicity proof.

Terms equipped with a computability structure
may be seen as a \emph{semantic domain} for interpretation.
To extract useful information from this domain,
we define two functions mutually recursively.
\[\text{Neutral form} \xrightarrow{\;\reflect\;}
\text{Computability} \xrightarrow{\;\reify\;}
\text{Normal form}\]
Reification takes a value in the semantic domain
and turns it into something concrete, i.e. a syntactic normal form.
Since normal forms are mutually inductively defined with neutral forms,
when defining reification we are naturally led to
simultaneously defining reflection.
Reflection and reification are also known as unquoting and quoting.

Recall that a computability structure for \(\Gamma \vdash t \of \Ans\)
is simply a normal form judgementally equal to \(t\).
Therefore we can define \(\reify_{\Gamma}^{\Ans}(u) = u\),
where we annotate the context and the type for clarity.
For the rest, we proceed by induction on the types.
\begin{align*}
\reify_{\Gamma}^{A \times B}(u) &= (\reify_{\Gamma}^{A}(\pi_1(u)), \reify_{\Gamma}^{B}(\pi_2(u))) \\
\reify_{\Gamma}^{\Unit}(u) &= ()\\
\reify_{\Gamma}^{A \to B}(u) &= u((\Gamma, x \Of A), \mathfrak{p}, x, \reflect_{\Gamma, x \Of A}^A(x))
\end{align*}
We invoked reflection in the function case, where
\(\mathfrak{p}\) is the obvious renaming from \(\Gamma, x \Of A\) to \(\Gamma\).
For reflection, we perform induction on the type.
For \(\Ans\), since a neutral form is already normal,
and a computability structure is simply a normal form,
we let \(\reflect_{\Gamma}^{\Ans}(n) = n\).
\begin{align*}
\reflect_{\Gamma}^{A \times B}(n) &=
    (\reflect_{\Gamma}^{A}(\pi_1(n)), \reflect_{\Gamma}^{B}(\pi_2(n)))\\
\reflect_{\Gamma}^{\Unit}(n) &= \atom{nil} \\
\reflect_{\Gamma}^{A \to B}(n) &= (\Theta, \sigma, a, u)
    \mapsto \reflect_{\Theta}^{B}(n[\sigma](\reify_{\Theta}^{A}(u)))
\end{align*}

Now, for every term \(\Gamma \vdash t \of A\),
\(\Sem{\Gamma \vdash t \of A}\) gives us a computability morphism.
We denote \(\vec x\) to be the identity substitution
\(\Gamma \vdash \vec x \of \Gamma\), with which we can instantiate
\[\semantics(t) = \Sem{\Gamma \vdash t \of A}_{\Gamma}(\vec x, \reflect_{\Gamma}^{\Gamma}(\vec x))\]
obtaining a normal form \(N = \reify_{\Gamma}^{A}(\semantics(t))\).
Here \(\reflect_{\Delta}^{\Gamma}(\vec a)\)
simply applies \(\reflect\) to each component.
We have defined the normalization function.
Finally a few straightforward induction proves the desired properties.
\end{proof}
\begin{remark}
Renamings are the most general class of substitutions we can allow here.
But we can also choose others.
For example, context containment \(\Gamma \supseteq \Delta\)
restricts \(\Delta\) to be \(\Gamma\) with some variables removed.
These are slightly harder to use in the dependent case.
\end{remark}

We may modularly add extra types and rules.
For example, we can add a case analysis principle to \(\Ans\),
making it a genuine Boolean type.
All the proofs can be extended respectively.
As a side note,
such a principle usually does not qualify as a full universal property.
This is because most languages do not have the rule
\[\inferrule{\Gamma, x \Of \Bool \vdash t \of A
\and \Gamma \vdash b \of \Bool}
{\Gamma \vdash \caseof(b, t[x/\true], t[x/\false]) = t[x/b] \of \Bool}\]
Of course one can add the rule in,
as has been done by \textcite{altenkirch:2004:nbe-bool}.
This type theory contains many unexpected judgemental equalities,
such as
\[x \of \Bool, f \of \Bool \to \Bool \vdash f(f(f(x))) = f(x).\]
And such principles cannot be extended to natural numbers
without breaking decidability.

\section{Simple Types, Categorically}

Our first step to sophistication is to consider
the category of \emph{contexts and substitutions} \(\mathcal T\).
Its objects are the contexts \(\Gamma, \Delta\),
and its morphisms are substitutions \(\Gamma \vdash \sigma \of \Delta\).
With tedious but standard induction over the syntax,
we can prove that substitutions are indeed associative and unital.
The empty context will serve as a terminal object of this category.

Recall that a \textbf{presheaf} over \(\mathcal T\)
is a functor \(\mathcal T\op \to \C{Set}\).
This is exactly the tool we need to handle substitutions automatically.
For example, for every context there is a collection of terms
(no matter whether they are considered up to \(\alpha\) equivalence
or \(\alpha\beta\eta\) equivalence),
which together form a presheaf,
with the functorial action given by term substitution.
\[\begin{matrix}
\Gamma & \xleftarrow{\;\;\sigma\;\;} & \Delta \\
t & \longmapsto & t[\sigma]
\end{matrix}\]
Or more granularly, for every type \(A\),
the terms of type \(A\) forms a presheaf \(\El(A)\),
since substitution preserves types.
Of course, for dependent types, substitution will also act on types.
So it is more convenient to put all the terms together in a single presheaf \(\Term\).
But for simple types, we will use \(\El(A)\) for simplicity.

To avoid an explosion of parentheses,
we write \(\El(\Gamma, A)\) for
the set of terms of type \(A\) in context \(\Gamma\).
So the presheaf \(\El(A)\) satisfies \(\El(A)(\Gamma) = \El(\Gamma, A)\).

\begin{idea}
Presheaves over the category of contexts and substitutions
allows us to use categorical language to handle substitution systematically.
\end{idea}

\subsection{Presheaves}

To wield the power of presheaves,
we first give a quick refreshening of the properties
of the category of presheaves \(\C{Psh}(\mathcal C)\).
We know that limits and colimits of presheaves are computed pointwise.
For example, for any two presheaves \(P, Q\),
\((P \times Q)(A) = P(A) \times Q(A)\),
and \((P + Q)(A) = P(A) + Q(A)\).
Here the operators on the right denote the product and disjoint union of sets.
It's easy to see that the universal properties are indeed satisfied.

Every category \(\mathcal T\) embeds into the presheaf category
\(\C{Psh}(\mathcal T)\) via the Yoneda embedding.
The Yoneda lemma states that \(\hom(\yo(A), X) \cong X(A)\)
naturally in \(A \of \mathcal T\) and \(X \of \C{Psh}(\mathcal T)\).
This allows for a very algebraic style of reasoning.
For example,
\begin{align*}
[\yo(B) \times \yo(C)](A)
&\cong\hom(\yo(A), \yo(B) \times \yo(C))\\
\explain{(product universal property)}\,
&\cong \hom(\yo(A), \yo(B)) \times \hom(\yo(A), \yo(C))\\
\explain{(Yoneda lemma)}\,
&\cong \hom(A, B) \times \hom(A, C)\\
&\cong \hom(A, B \times C)\\
\explain{(definition of \(\yo\))}\,
&= \yo(B \times C)(A)
\end{align*}
shows that \(\yo\) preserves any products (or more generally, any limits)
that exist. Similarly,
\begin{align*}
[\yo(B) \to \yo(C)](A)
&\cong \hom(\yo(A), \yo(B) \to \yo(C)) \\
\explain{(definition of exponentials)}\,
&\cong \hom(\yo(A) \times \yo(B), \yo(C)) \\
\explain{(\(\yo\) preserves products)}\,
&\cong \hom(\yo(A \times B), \yo(C)) \\
\explain{(Yoneda lemma)}\,
&\cong \hom(A \times B, C)\\
\explain{(definition of exponentials)}\,
&\cong \hom(A, B\to C)\\
&= \yo(B \to C)(A)
\end{align*}
shows that \(\yo\) preserves exponentials.
We can even compute a concrete representation of
exponential objects in the presheaf category.
Given two presheafs \(P, Q\),
\[
(P \to Q)(A) \cong \hom(\yo(A), P \to Q) \cong \hom(\yo(A) \times P, Q).
\]
In light of this identity, we can simply \emph{define} the exponential
\(P \to Q\) to be the presheaf \(\hom(\yo(-) \times P, Q)\).
More explicitly, an element of \((P \to Q)(A)\)
is a function \(f\) whose domain consists of triples \((B, \sigma, p)\),
where \(B\) is an object, \(\sigma \of B \to A\), and \(p \in P(B)\).
\(f\) maps \((B, \sigma, p)\) to an element of \(Q(B)\),
such that certain naturality squares commute.

Did that feel familiar?
In fact, this captures the essence of the Kripke construction!
Such a mysterious construction essentially comes from
the exponential object in presheaf categories.
We did not come up with anything clever to construct the exponential object.
It simply rolled out of a one-line computation!

\begin{idea}
Kripke function spaces can be naturally calculated
from the categorical definition of exponential objects in presheaf categories.
Other complicated constructions may be similarly obtained.
\end{idea}

\subsection{Exploring the syntax with presheaves}

Recall that \(\El(A)\) is the presheaf of terms of \(A\),
considered up to \(\beta\eta\) equality.
We have two maps
\[\El(A \times B)
\xrightleftharpoons[(-,-)]{\pi_1, \pi_2}
\El(A) \times \El(B)\]
They are natural transformations because substitutions commute with term formers.
The \(\beta\) and \(\eta\) laws state that they are mutual inverses.
Thus we have \(\El(A \times B) \cong \El(A) \times \El(B)\).
Similarly, \(\El(\Unit) \cong 1\),
where the latter \(1\) refers to the terminal presheaf \(\Gamma \mapsto \{\star\}\).
For \(\Ans\), since it is not determined by a universal property,
we don't have such an isomorphism.
Instead, we have two morphisms \(1 \rightrightarrows \El(\Ans)\)
representing \(\yes\) and \(\no\), respectively.
Note that this also automatically takes care of the substitutions
\(\yes[\sigma] = \yes\) etc.

What about function types?
Function application introduces a natural transformation
from \(\El(A\to B) \times \El(A)\) to \(\El(B)\),
which is equivalent to a map
\(\El(A \to B) \longrightarrow [\El(A) \to \El(B)]\).%
\footnote{I use long arrows for visual distinctiveness only.
They mean the exact same things as the short arrows \(\to\).}
On the other hand, we can compute the \(\hom\)-presheaf
using the identities proved above.
\[[\El(A) \to \El(B)](\Gamma) = \hom(\yo(\Gamma) \times \El(A), \El(B)).\]
The value of the presheaf \(\yo(\Gamma) \times \El(A)\)
at a context \(\Delta\) consists of pairs \((\sigma, a)\)
where \(\Delta \vdash \sigma \of \Gamma\) and \(\Delta \vdash a \of A\).
Combining these, we get exactly a substitution
\((\sigma, a) \in \hom(\Delta, \Gamma \cdot A)\),
where \(\Gamma \cdot A\) is the context extension of \(\Gamma\) by \(A\).%
\footnote{With named variables, we write \(\Gamma, x \Of A\).
But since names don't really matter,
and using commas to denote an operation is confusing,
we use \(\Gamma \cdot A\).}
We may continue the computation
\begin{align*}
[\El(A) \to \El(B)](\Gamma)
&= \hom(\yo(\Gamma) \times \El(A), \El(B)) \\
&\cong \hom(\yo(\Gamma\cdot A), \El(B)) \\
\explain{(Yoneda lemma)} \quad &\cong \El(B)(\Gamma \cdot A) \\
\explain{(by definition)} \quad &= \El(\Gamma \cdot A, B)
\end{align*}
which is exactly the set of \(t\) such that \(\Gamma, x \Of A \vdash t \of B\).
The \(\lambda\) abstraction operation produces,
for each of these terms,
a term \(\Gamma \vdash \lambda x\bind t \of A \to B\).
Therefore it induces a map
\([\El(A) \to \El(B)] \longrightarrow \El(A \to B)\),
so we may write
\[\El(A \to B) \xrightleftharpoons[\lambda]{-(-)} \El(A) \to \El(B).\]
\(\beta\) and \(\eta\) laws state that they are mutual inverses.
This looks awfully like a technique for encoding syntax bindings
called \emph{higher-order abstract syntax}.
We will come to this later.

\begin{idea}
As presheaves over the syntactic category,
exponential objects serve as the binding operator,
enabling a form of higher-order abstract syntax.
\end{idea}

\subsection{Category with families}

We expect semantics to have matching structures with syntax.
Since in the syntax we have characteristic isomorphisms
\(\El(A \times B) \cong \El(A) \times \El(B)\) and
\(\El(A \to B) \cong \El(A) \to \El(B)\),
they should be present in the semantics.

\begin{definition}\label{def:model:scwf}
A \textbf{simple category with families} consists of the following data:
\begin{itemize}
\item A category \(\mathcal C\), a set \(\Type\), and
a map \(\El \of \Type \to \C{Psh}(\mathcal C)\)
assigning a presheaf over \(\mathcal C\) for each element of \(\Type\),
\item A terminal object \(1\) of \(\mathcal C\),
and a binary operation taking \(\Gamma \of \mathcal C\)
and \(A \in \Type\) to an object \(\Gamma \cdot A \of \mathcal C\),
equipped with an isomorphism
\[\yo(\Gamma \cdot A) \cong \yo(\Gamma) \times \El(A),\]
\end{itemize}
A \textbf{model} of simply typed lambda calculus
further has chosen elements \(\Unit, \Ans\)
and binary operations \((\times), (\to)\) in \(\Type\),
with natural isomorphisms
\begin{align*}
\El(\Unit) &\cong 1 \\
\El(A \times B) &\cong \El(A) \times \El(B)\\
\El(A \to B) &\cong \El(A) \to \El(B)
\end{align*}
and two morphisms \(1 \rightrightarrows \El(\Ans)\).
For convenience we give names to these morphisms.
\begin{code}
\>[0]\(\synct{unit} \of 1 \to \El(\Unit)\)\\
\>[0]\(\synct{pair} \of \El(A) \times \El(B) \to \El(A \times B)\)\\
\>[0]\(\synct{fst} \of \El(A \times B) \to \El(A)\)\\
\>[0]\(\synct{snd} \of \El(A \times B) \to \El(B)\)\\
\>[0]\(\synct{lam} \of (\El(A) \to \El(B)) \to \El(A \to B)\)\\
\>[0]\(\synct{app} \of \El(A \to B) \times \El(A) \to \El(B)\)\\
\>[0]\(\synct{yes} \of 1 \to \El(\Ans)\)\\
\>[0]\(\synct{no} \of 1 \to \El(\Ans)\)
\end{code}
\end{definition}
\begin{remark}
Although we used similar notations, it is vital to keep in mind that
our base category \(\mathcal C\) is not necessarily the same as the syntactic one,
\(\Type\) not necessarily the set of syntactic types,
and \(\El\) not necessarily a presheaf of syntactic terms.
We will sometimes refer to them as \emph{semantic} gadgets for emphasis.
For example, an object of \(\mathcal C\) is called a \emph{semantic context},
and an element of \(\Type\) a \emph{semantic type}.
\end{remark}

Although we used presheaves to bookkeep naturality conditions,
the definition is still essentially \emph{algebraic},
in the sense that it is governed by operations and equations between them.
Thus we have an automatic notion of morphisms between models,
in line-by-line correspondence of the definition of models.

\begin{definition}\label{def:morphisms:scwf}
Given two models \(\mathcal C, \mathcal D\)
a \textbf{morphism} between them consists of the following data:
\begin{itemize}
\item A functor \(F \of \mathcal C \to \mathcal D\),
a map \(\Type_{\mathcal C} \to \Type_{\mathcal D}\)
which we also write \(F\) by abuse of notation,
and a family of natural transformations
\[F_A \of \El_{\mathcal C}(\Gamma, A) \to \El_{\mathcal D}(F(\Gamma), F(A))\]
natural in \(\Gamma \of \mathcal C\) for each \(A \in \Type_{\mathcal C}\).
\item Equalities for each operation
\begin{align*}
F(1_{\mathcal C}) &= 1_{\mathcal D}
\quad \mathrlap{\explain{(chosen terminal object)}} \\
F(\Gamma \cdot A) &= F(\Gamma) \cdot F(A) \\
F(\Unit) &= \Unit \\
F(\Ans) &= \Ans \\
F(A \times B) &= F(A) \times F(B) \\
F(A \to B) &= F(A) \to F(B),
\end{align*}
\item Equalities for each morphism,
requiring \(F\) respect \(\synct{yes}, \synct{no}\) and all the isomorphisms.
\end{itemize}
\end{definition}
\begin{remark}
The reader seasoned in the \(n\)-PoV~\cite{nlab:2023:npov}
may cringe at the sight of strict equalities between objects in a category.
Indeed there is a notion of \emph{pseudomorphisms}
that requires the relevant comparison maps to be isomorphisms,
instead of equalities.
The reader can refer to \textcite{castellan:2020:cwf} for more discussion,
or \textcite{ahrens:2018:cwf-univalent} for a univalent point of view.
\end{remark}

How can we confirm that we made the correct definition of models?
Traditionally, this question is answered by proving
\emph{soundness} and \emph{completeness} lemmas.
With our categorical language, they take on new forms.

\begin{lemma}\label{lemma:syntactic-cwf}
The syntax forms a model \(\mathcal T\)
called the \textbf{syntactic model}.
\end{lemma}
\begin{proof}
Trivial by construction.
\end{proof}
\begin{remark}
This corresponds to the completeness in traditional formal systems.
Since syntax itself forms a model,
a property of all models would then tautologically be a property of syntax too.
We have deliberately defined models so that the proof is trivial.
But in cases like first-order logic,
syntactic models are constructed in a slightly non-trivial fashion
through what is known as \emph{Herbrand structures}.
\end{remark}

\begin{lemma}\label{lemma:initial-cwf}
The syntactic model is \emph{initial}.
In other words, for any other model \(\mathcal M\),
there is a unique morphism \(F \of \mathcal T \to \mathcal M\).
\end{lemma}
\begin{proof}
For the map on \(\Type\), we can perform induction on the syntactic types.
In the base case,
\(F(\Unit) = \Unit\) and \(F(\Ans) = \Ans\) are uniquely specified.
If the image of \(A\) and \(B\) under \(F\) are determined,
then \(F(A \times B) = F(A) \times F(B)\) and \(F(A \to B) = F(A) \to F(B)\)
determines the mapping of \(F\) on product and function types.
Therefore \(F\) is completely fixed on the types.

For the map on objects,
all syntactic contexts are of the form
\((((1 \cdot A_1) \cdot A_2) \cdot \cdots) \cdot A_n\)
where \(1\) is the empty context.
\(F(1) = 1\) is fixed because
\(1\) is the chosen terminal object.
And since we already proved that \(F(A_i)\) are all fixed,
by \(F(\Gamma \cdot A) = F(\Gamma) \cdot F(A)\)
we see that \(F\) is uniquely determined on the objects of \(\mathcal T\).

For morphisms we perform induction on the syntactic terms.
Since \(F\) is required to preserve the relevant morphisms,
each term former gets uniquely determined.
And since we consider terms (and thus substitutions) up to jugdemental equality,
we need to check that equal substitutions are mapped to equal morphisms.
This is true because of the \(\beta, \eta\) laws.
Similarly for the action of \(F_A\) for each type \(A\).
\end{proof}

\begin{remark}
This is the categorical codification of induction on syntax,
and corresponds to the soundness theorem.
It packages up and allows us to completely do away with
the quirks of syntax and binding trees.
In fact, for most purposes in type theory,
we can take this as a \emph{definition} of syntax.
This radical step allows us to avoid dealing with
boring substitution lemmas altogether.
\end{remark}
\begin{idea}
Initiality expresses induction on syntax in a more elegant
and implementation-agnostic way.
\end{idea}

\begin{theorem}[Equational consistency]
\(\Gamma \nvdash \yes = \no \of \Ans\).
\end{theorem}
\begin{proof}
We construct a model over the category \(\C{Set}\).
Take \(\Type\) to be all the sets, and \(\El(\Gamma, A) = \hom(\Gamma, A)\).
Let \(\Ans = \{\atom{t}, \atom{f}\}\) be a two-element set,
and \(1 \rightrightarrows \El(\Ans)\) select the two constant functions.
The rest of the structure is entirely straightforward.
Context extension is defined by \(\Gamma \cdot A = \Gamma \times A\)
using the Cartesian product of sets.
Since we defined \(\El(A) = \yo(A)\)
(which only makes sense because we defined \emph{both}
the objects and types as sets),
we have \(\El(A \times B) \cong \El(A) \times \El(B)\)
and \(\El(A \to B) \cong \El(A) \to \El(B)\).
This forms a model.

By \Cref{lemma:initial-cwf}, we have a unique morphism
\(\Sem{-} \of \mathcal T \to \C{Set}\).
Since \(\Sem{\yes} \ne \Sem{\no} \in \El(\Sem{\Gamma}, \Sem{\Ans})\),
we conclude that \(\yes \ne \no\).
\end{proof}
\begin{remark}
Since models are algebraic structures,
this proof is essentially in the same spirit as
proving \(xy \ne yx\) in the free group \(F_2 = \langle x, y \rangle\),
by exhibiting an explicit non-commutative group \(G\),
obtaining a homomorphism \(F_2 \to G\),
seeing that the images of \(xy\) and \(yx\) are different,
and concluding that \(xy = yx\) cannot hold.
Indeed, our initiality lemma~\ref{lemma:initial-cwf}
exactly states that the syntactic model is the free model
(over no generators).

There is actually much flexibility in our construction of models.
For example we can take \(\Type\) to be the set of syntactic types instead.
This is also similar to the group-theoretic example,
because there are loads of non-commutative groups
and any one of them can do.
\end{remark}

\begin{remark}
Taking the set of all sets unfortunately leads to size-related paradoxes.
Fortunately this is easy to circumvent with standard set-theoretic techniques.
If the reader would like to keep it simple, assume there is a set of all sets.
But keep in mind that size issues are indeed important in category theory.%
\cite{shulman:2008:set}

Working in ZFC for example, take any limit ordinal \(\alpha\).
(\(\alpha = \omega\) is good enough).
Recall the von Neumann universe is constructed via
\(V_0 = \varnothing\), \(V_{\alpha + 1} = \mathcal{P}(V_\alpha)\),
and \(V_{\alpha} = \bigcup_{\lambda < \alpha} V_\lambda\) for limit ordinals.
If \(X, Y \in V_{\lambda}\),
then \((X \times Y) \in V_{\lambda + 2}\)
and \((X \to Y) \in V_{\lambda + 3}\).
Therefore if \(\alpha\) is limit, then
\(V_\alpha\) will be closed under products and exponentials.
So we can use the corresponding small category \(\C{Set}_\alpha\).
Now taking \(\Type = V_\alpha\) is no longer problematic.
Of course, if our type theory has natural numbers
or any other way to get bigger types,
we would need larger ordinals.
\end{remark}

\subsection{Gluing}\label{subsec:scwf:gluing}

We can now have a partial reveal of
the categorical nature of computability structures.
We will set up a model \(\mathcal G\) of computability structures,
with a morphism \(\pi \of \mathcal G \to \mathcal T\) to keep track of
which term the computability structure corresponds to.
By \Cref{lemma:initial-cwf}, we have a morphism
\(\Sem{-} \of \mathcal T \to \mathcal G\),
assigning a computability structure \(\Sem{t}\) to every term \(t\).

Composing we have \(\pi \comp \Sem{-} \of \mathcal T \to \mathcal T\).
But by initiality there can only be one morphism from \(\mathcal T\) to any model.
And since the identity morphism is \(\mathcal T \to \mathcal T\),
we conclude that \(\pi \comp \Sem{-} = \textrm{Id}\).
This is a very useful annex to initiality.
A model together with a morphism back to the initial model is
known as \textbf{displayed models} in the literature.~\cite{ahrens:2017:displayed}
In addition to an interpretation,
it helps keep track of the relation of
each piece syntax and its interpretation,
in turn ensuring that \(\Sem{t}\) really is a computability structure for \(t\),
and not for any other term.

There is a systematic way to construct displayed models called \emph{gluing}.
It has many manifestations, and the name comes from one of its geometric forms.
Such a technique in type theory is worked out by \textcite{kaposi:2019:gluing}.
We give a slightly different presentation here.

The material needed for gluing is two simple CwFs
\(\mathcal C, \mathcal D\) and a sort of morphism
\(F \of \mathcal C \to \mathcal D\).
However, we do not need \(F\) to strictly preserve every structure,
and most of the examples we will see do not.
It only need to respect context extensions up to isomorphism.
\begin{definition}
A \textbf{pseudomorphism} \(F \of \mathcal C \to \mathcal D\)
is a functor \(\mathcal C \to \mathcal D\),
a map \(\Type_{\mathcal C} \to \Type_{\mathcal D}\),
and a family of natural transformations
\(\El_{\mathcal C}(\Gamma, A) \to \El_{\mathcal D}(F(\Gamma), F(A))\),
such that the canonical morphisms
\(F(1_{\mathcal C}) \to 1_{\mathcal D}\) and
\(F(\Gamma \cdot A) \to F(\Gamma) \cdot F(A)\)
are isomorphisms.
\end{definition}
When \(\mathcal C, \mathcal D\) are actually models
of simply typed lambda calculus,
\(F\) does not necessarily preserve the type structures.
In this case, following \textcite[5.5.2\textasteriskcentered3]{sterling:2021:thesis},
we call it a \emph{Henkin morphism} to disambiguate.
We have a definition of
\emph{glued} simple CwF along a pseudomorphism
\(F \of \mathcal C \to \mathcal D\)
in components:
\begin{itemize}
\item The objects are triples
\(\Gamma = (\Gamma^\sem, \Gamma^\downarrow, \Gamma^\syn)\)
where \(\Gamma^\sem \of \mathcal D\),
\(\Gamma^\syn \of \mathcal C\) and
\(\Gamma^\downarrow \of \Gamma^\sem \to F(\Gamma^\syn)\).
\item The morphisms are commutative squares
\[\begin{tikzcd}
    {\Gamma^\sem} & {\Delta^\sem} \\
    {F(\Gamma^\syn)} & {F(\Delta^\syn)}
    \arrow[from=1-1, to=2-1]
    \arrow[from=1-2, to=2-2]
    \arrow["{F(\sigma^\syn)}", from=2-1, to=2-2]
    \arrow["{\sigma^\sem}", from=1-1, to=1-2]
\end{tikzcd}\]
In other words, the underlying category of the model
is the \emph{comma category} \(\textrm{Id}_{\mathcal D} \CommaCat F\).
\item The types are triples \((A^\sem, A^\downarrow, A^\syn)\)
where \(A^\downarrow\) is a natural transformation
\(\El_{\mathcal D}(A^\sem) \to \El_{\mathcal D}(F(A^\syn))\).
\item \(\El(\Gamma, A)\) consists of pairs \((a^\sem, a^\syn)\)
such that \(A^\downarrow(a^\sem) = F(a^\syn)[\Gamma^\downarrow]\).
This agrees with morphisms \(\hom(\Gamma, \Delta)\)
if \(\Delta\) is a single-type context.
\item \(\Gamma \cdot A\) is the triple
\((\Gamma^\sem \cdot A^\sem, \sigma, \Gamma^\syn \cdot A^\syn)\)
where \(\sigma\) is the composite
\[\Gamma^\sem \cdot A^\sem \longrightarrow
F(\Gamma^\syn) \cdot F(A^\syn) \xrightarrow{\;\cong\;}
F(\Gamma^\syn \cdot A^\syn).\]
One can verify that \(\yo(\Gamma \cdot A) \cong \yo(\Gamma) \times \El(A)\)
by routine calculation.
\end{itemize}

We have an obvious projection
\((\Gamma^\sem, \Gamma^\downarrow, \Gamma^\syn) \mapsto \Gamma^\syn\),
which is a (strict) morphism by construction.
Therefore if we take \(\mathcal C\) to be the syntactic model,
\(\mathcal D\) can be viewed as the raw computability data,
and gluing along \(F\) constructs the ``tracking system''.

\begin{idea}
Displayed models enhance the power of initiality,
upgrading it from the recursion principle to the full dependent elimination principle.
Gluing is a systematic way to produce such models.
\end{idea}

\begin{theorem}[Canonicity]\label{theorem:canonicity:categorical}
For every term \(\vdash t \of \Ans\),
either \(\vdash t = \yes \of \Ans\) or \(\vdash t = \no \of \Ans\) is derivable.
\end{theorem}
\begin{proof}
We can define a Henkin morphism
\(\GSect \of \mathcal T \to \C{Set}\)
by \(\GSect(\Delta) = \hom(1, \Delta)\) on the underlying category,
\(\GSect(A) = \El(1, A)\).
We have obvious maps natural in \(\Delta\)
\[\El_{\mathcal T}(\Delta, A) \longrightarrow
\overbrace{\El_{\C{Set}}(\GSect(\Delta), \GSect(A))}
^{\hom(1, \Delta) \to \El(1, A)}.\]
This forms a Henkin morphism because
\(\GSect(1) = \hom(1, 1) \cong 1\) and
\(\GSect(\Gamma \cdot A) = \yo(\Gamma \cdot A)(1)
\cong \yo(\Gamma)(1) \times \El(A)(1) = \GSect(\Gamma) \times \El(\GSect(A))\),
so it does preserve structure up to isomorphism.
\(\GSect\) is called the \emph{closed terms} Henkin morphism
because, well, it selects the closed terms.
A more geometric word is \emph{global sections} functor.

We consider the glued simple CwF
\(\mathcal G = \textrm{Id}_{\C{Set}} \CommaCat \GSect\).
As an exercise for the reader, unfold the definitions
and try to recover the constructions in the naïve proof.
For example, the types of this model are triples,
\(A^\sem\) is the set of computability structures,
\(A^\syn\) is the actual type, and
\(A^\downarrow \of \El_{\C{Set}}(A^\sem) \to \El_{\C{Set}}(\GSect(A^\syn))\)
maps the computability structures to their corresponding closed terms.

We need to reconstruct the type structures to obtain a model.

\begin{itemize}
\item For the unit type, \(\Unit^\syn = \Unit_{\mathcal T}\),
\(\Unit^\sem = \{\atom{nil}\}\), and \(\Unit^\downarrow\) is the unique morphism.
\item For products we set \((A \times B)^\syn = A^\syn \times B^\syn\),
\((A \times B)^\sem = A^\sem \times B^\sem\).
\((A \times B)^\downarrow\) will be obvious after unfolding the definition.
\item We set \((A \to B)^\syn = A^\syn \to B^\syn\), but
\begin{equation}
(A \to B)^\sem = \left\{(f^\syn, f^\sem) ~\middle|~
\begin{gathered}
f^\syn \in \El_{\mathcal T}(1, A^\syn \to B^\syn) \\
f^\sem \in A^\sem \to B^\sem \\
B^\downarrow \comp f^\sem = f^\syn \comp A^\downarrow
\end{gathered}\right\}
\tag{\(\star\)}
\label{eq:function:scwf}
\end{equation}
which can be seen as the internalization (as in the name ``internal \(\hom\)'')
of the commuting squares condition.
\((A \to B)^\downarrow\) projects to the first component.
This construction is specific to \(\C{Set}\),
so to glue any other model we need to come up with something else.
\item Finally for \(\Ans\) we set \(\Ans^\syn = \Ans\),
\(\Ans^\sem = \{\atom{t}, \atom{f}\}\),
sending \(\atom{t}\) to \(\yes\) and \(\atom{f}\) to \(\no\).
\end{itemize}
This defines a model.
Therefore by the aforementioned argument,
we have an interpretation \(\Sem{-}\) that
in particular maps each closed term \(\vdash t \of \Ans\) to
an element \(\Sem{t} \in \El_{\mathcal G}(1, \Ans)\),
which is a pair \((\Sem{t}^\sem, \Sem{t}^\syn)\) satisfying an equation.
We already argued that \(\Sem{t}^\syn = t\), ensuring that
the computability structure indeed corresponds to the term \(t\).
\(\Sem{t}^\sem\) is a map from the singleton set to \(\{\atom{t}, \atom{f}\}\).
Whichever it is, the satisfied equation ensures that \(t\) would then be
\(\yes\) or \(\no\).
\end{proof}

There is in fact quite a lot of detail that I did not write down here.
They are mostly functoriality and naturality proof obligations,
which are very tedious to fill.
Of course, the construction of most of them is unique,
because they have been fixed by isomorphisms in our definition.
Such an avalanche of technical details is what synthetic Tait computability sets out to solve.
But before we do that, let's do normalization in this language first.

\subsection{Normalization}

For the normalization proof,
we definitely can't use the functor \(\GSect(\Gamma) = \hom(1, \Gamma)\) anymore,
since we want to deal with terms in all contexts.
A natural choice would be instead using
\(\yo(\Gamma) = \hom(-, \Gamma)\), a functor \(\mathcal T\op \to \C{Psh}(\mathcal T)\),
because it includes the information at all contexts.
However, as we mentioned, normal forms and neutral forms
are not stable under arbitrary substitution.
Therefore we need to restrict to a subcategory \(\mathcal A\)
of contexts and renamings.
We have an obvious functor \(\rho \of \mathcal A \hookrightarrow \mathcal T\).
There is the \emph{restricted Yoneda} functor
\(\Nerve \of \mathcal T \to \C{Psh}(\mathcal A)\),
defined as
\[\Nerve(\Gamma) = \hom(\rho(-), \Gamma).\]
We can now define our new
\(\mathcal G = \textrm{Id}_{\C{Psh}(\mathcal A)} \CommaCat \Nerve\).
and proceed to prove that it is a model.
\begin{remark}
If we replace \(\mathcal A\) with the category \(1\),
and let \(\rho\) point to the terminal object,
we get back \(\mathcal G\) in the canonicity proof.
\end{remark}

\(\C{Psh}(\mathcal A)\) behaves extremely like \(\C{Set}\),
having products, pullbacks, exponentials, etc.
In fact all the constructs required in the canonicity proof can be carried out.
The most notable one is the construction of function types.
The set comprehension in \cref{eq:function:scwf} can be re-expressed as
\[\begin{tikzcd}
\bullet & {A^\sem \to B^\sem} \\
{\El(1, A^\syn \to B^\syn)} & {A^\sem \to \El(1, A^\syn)}
\arrow[from=1-1, to=2-1]
\arrow[""{name=0, anchor=center, inner sep=0}, from=2-1, to=2-2]
\arrow[from=1-2, to=2-2]
\arrow[from=1-1, to=1-2]
\arrow["\lrcorner"{anchor=center, pos=0.1}, draw=none, from=1-1, to=0]
\end{tikzcd}\]
Now we can replace \(\GSect(A^\syn) = \El(1, A^\syn)\) with our \(\Nerve\).
This precisely reconstructs the Kripke logical relation
as appeared in our naïve proof.
The exponential \(A^\sem \to B^\sem\) produces the weird renaming dance,
and the pullback fetches the subset that correctly tracks the terms.
There is not much to add here, so we omit the proof.

So far, the main advantage of categorical language is that
it makes a large part of the construction unique,
and therefore once we give a solution we can be sure that it is the right one.
We no longer need to proceed by trial and error to find
``the thing that lets induction go through''.
To quote \textcite[4.2\textasteriskcentered12]{sterling:2021:thesis}:
\begin{quotation}
For this reason, the usual emphasis on careful constructions of logical relations
for (e.g.) product types or function types
that pervades the literature must be regarded as essentially misguided,
an artifact of the now-eclipsed era in which
the universal properties of type theoretic connectives were either omitted
or hidden underneath inessential syntactical matters such as raw terms.
\end{quotation}
However, manually filling in the coherence proofs is very tedious,
and in fact most of them should simply be automatic.
On the other hand, we cannot handwave the problems away,
because some constructions such as universes and cubical Kan operations
have a real danger that seemingly natural constructions
do not in fact satisfy naturality.

Such a dilemma can be solved by going into an \emph{internal language} of
the category under question.
In such a language, by virtue of being able to write down an object
we know that it will satisfy all the required conditions.

\section{Simple Types, Synthetically}

Topoi are categories equipped with rich structures.
(Let's not worry about its precise definition for now.)
So rich, in fact, that they can support a sort of
internal logic where the objects are treated like \emph{sets}.
We can work in such a logic axiomatically,
and later instantiate a specific topos
to obtain complex theorems using simple language.
This style of reasoning is often used
in the active field of \emph{synthetic mathematics}.

The language differs from our usual language in that
\emph{excluded middle} is taken away, i.e.
it is not necessarily true that \(\forall p, p \lor \neg p\).
Of course, in some topoi such as \(\C{Set}\) and \(\C{FinSet}\) it holds.
And in other topoi such as \(\C{Set}^{\to}\),
\(\neg(\forall p, p \lor \neg p)\) is true.
Note that this is different from \(\exists p, \neg(p \lor \neg p)\),
which is false in every topos.

We will see a plethora of examples of important topoi.
In particular, all (small) presheaf categories are topoi,
and \emph{Artin gluings} of topoi --- as will be introduced later ---
are also topoi.
We shall simply introduce as needed the structures interpreting
the internal language.

\subsection{Upgrading to topoi}

The syntactic category \(\mathcal T\),
or generally the underlying category of a model is not a topos.
But we are effectively working in a topos
--- the presheaf topos over them,
because \(\El(A)\) are all presheaves.
This is too restrictive because we want to use non-presheaf topoi too.
We will temporarily upgrade to an \emph{ad hoc} notion of model
tailored for use in topoi,
because the necessary framework to make this systematic and motivated
is best introduced in the dependent case.
We also prepare some lemmas to help us upgrade.

\begin{definition}\label{def:model:topos}
A \textbf{STLC structure} over a
Cartesian closed category \(\mathcal E\) consists of the following data:
\begin{itemize}
\item A set \(\Type\) and a map \(\El \of \Type \to \mathcal E\),
\item Chosen elements \(\Unit, \Ans\)
and binary operations \((\times), (\to)\) in \(\Type\),
with natural isomorphisms
\begin{align*}
\El(\Unit) &\cong 1 \\
\El(A \times B) &\cong \El(A) \times \El(B)\\
\El(A \to B) &\cong \El(A) \to \El(B)
\end{align*}
and two morphisms \(1 \rightrightarrows \El(\Ans)\).
\end{itemize}
\end{definition}
\begin{lemma}
Every model of simply typed lambda calculus
corresponds to a STLC structure over some category.
\end{lemma}
\begin{proof}
Take \(\mathcal E = \C{Psh}(\mathcal C)\).
\end{proof}
\begin{remark}
Much of the structure is simplified because
we originally keep track of whether an object is in \(\mathcal C\),
and we require the context extension operation to stay in \(\mathcal C\),
instead of being in \(\C{Psh}(\mathcal C)\).
Now we let go of this.
\end{remark}
\begin{definition}
Given \(\mathcal E, \mathcal F\) equipped with STLC structures,
a morphism between them is
a Cartesian closed functor \(F \of \mathcal E \to \mathcal F\)
and a map \(\Type_{\mathcal E} \to \Type_{\mathcal F}\)
such that
\begin{itemize}
\item \(\El_{\mathcal F}(F(A)) = F(\El_{\mathcal E}(A))\).
\item \(F(\Unit) = \Unit\),
\(F(A \times B) = F(A) \times F(B)\),
\(F(A \to B) = F(A) \to F(B)\) and
\(F(\Ans) = \Ans\).
\item The functor \(F\) maps the isomorphism \(\El(\Unit) \cong 1\) in
\(\mathcal E\) to the isomorphism in \(\mathcal F\),
similarly for \(A \times B\), \(A \to B\).
And \(F(\synct{yes}) = \synct{yes}, F(\synct{no}) = \synct{no}\).
Note that these equalities make sense because the previous conditions
ensured that the domains and codomains are equal.
\end{itemize}
\end{definition}

This upgrading doesn't affect our \(\pi \comp \Sem{-}\) argument
in \Cref{subsec:scwf:gluing}.
\begin{lemma}\label{lemma:quasiprojectivity}
For an arbitrary morphism of STLC structures
\(\pi \of \mathcal G \to \C{Psh}(\mathcal T)\),
we have
\begin{itemize}
\item A functor \(\Sem{-} \of \mathcal T \to \mathcal G\)
with an isomorphism \(\pi \comp \Sem{-} \cong \yo\).
\item A map on types \(\Sem{-} \of \Type_{\mathcal T} \to \Type_{\mathcal G}\)
with \(\pi \comp \Sem{-} = \textrm{Id}\).
\item The empty context has \(\Sem{1} \cong 1\), and context extension has
\(\Sem{\Gamma \cdot A} \cong \Sem{\Gamma} \times \El_{\mathcal G}(\Sem{A})\)
whose image under \(\pi\) is the isomorphism \(\yo(\Gamma \cdot A) \cong \yo(\Gamma) \times \El_{\mathcal T}(A)\).
\item A family of natural transformations on terms
\(\Sem{-}_A \of \El_{\mathcal T}(A) \to \El_{\mathcal G}(\Sem{A})\)
such that \(\pi_A \comp \Sem{-}_A \cong \textrm{Id}\).
\end{itemize}
\end{lemma}
\begin{remark}
This is called \emph{quasi-projectivity} in \cite{gratzer:20??:thesis}.
In some sense this records the partial power of initiality that
still remains after moving from simple CwFs to STLC structures.
We will live with that because our notion of STLC structures
is somewhat \emph{ad hoc}.
\end{remark}
\begin{proof}
Straightforward using the techniques from last section.
We construct a model \(\mathcal D\) whose semantic objects are triples
\((\Gamma^\sem, \Gamma^{\cong}, \Gamma^\syn)\) where
\(\Gamma^\sem \of \mathcal G\), \(\Gamma^\syn \of \mathcal T\),
and \(\Gamma^{\cong} \of \pi(\Gamma^\sem) \cong \yo(\Gamma^\syn)\).
The types \(\Type_{\mathcal D} = \Type_{\mathcal G}\), with
\[\El_{\mathcal D}(\Gamma, A) =
\left\{(a^\sem, a^\syn) \,\middle|\,
\begin{gathered}
a^\sem \in \hom(\Gamma^\sem, \El_{\mathcal G}(A)) \\
a^\syn \in \El_{\mathcal T}(\Gamma^\syn, \pi(A)) \\
\pi(a^\sem) \sim a^\syn
\end{gathered}
\right\}\]
where \(u \sim v\) means \(u\) is mapped to \(v\)
when suitably acted on with \(\Gamma^{\cong}\) and the Yoneda isomorphism.
\end{proof}

\begin{idea}
We upgrade our syntactic category to a topos,
so that we can access more structure.
The core power of initiality still remains after such an upgrade.
\end{idea}

\subsection{An Internal Language of Gluing}

The gluing construction with topoi is particularly well studied.
It takes two topoi \(\mathcal E, \mathcal F\) and
a finitely continuous functor \(F \of \mathcal E \to \mathcal F\).
As a category, it is simply the comma category
\(\textrm{Id}_{\mathcal F} \CommaCat F\).
It is proved to be another topos in
\cite[Example A2.1.12]{johnstone:2008:elephant}.
Again, let's not worry about
what a topos is actually defined to be for the moment,
but topoi are all Cartesian closed,
so they have enough structure for the simple types.

The glued topos \(\mathcal G\) comes naturally with two projection functors
\(\pi_\syn \of \mathcal G \to \mathcal E\)
and \(\pi_\sem \of \mathcal G \to \mathcal F\).
The first is a Cartesian closed functor as we shall see later,
and the second is kind of janky.
When \(\mathcal E = \C{Psh}(\mathcal T)\),
we can use \Cref{lemma:quasiprojectivity} with \(\mathcal G\) and \(\pi_\syn\),
obtaining a computability structure for each term.
But recall that the hard part here is
to fill in the STLC structure for \(\mathcal G\),
and it was a lot of work in the categorical proof.

Synthetic Tait computability cuts through the red tape
by working in the internal language of \(\mathcal G\).
Topoi admit an internal language of higher order logic,
or extensional dependent type theory with a proposition type.
These are inter-translatable~\cite{jacobs:1993:translation}.
Dependent type theory is easier to use,
but the connection of this internal language to the external world
is slightly harder to explain.
The connection between higher order logic and topos theory
is far more well studied, for example by \textcite{johnstone:2008:elephant}.

Building on this, we can add postulates that create additional structures,
as long as we can interpret them back to \(\mathcal G\).
Then, we can work purely inside the internal language,
giving clean and concise definitions and proofs,
which are then externalized to obtain our final result.
We give a rough intuition of the vocabulary of the internal language,
and what external construction it corresponds to.

A type (or a set) in our logic corresponds to an object in \(\mathcal G\).
We can perform the usual operations on types.
Binary product is interpreted as categorical product in \(\mathcal G\),
which is constructed as the middle column in the diagram.
\[\begin{tikzcd}
    {X^\sem} & {X^\sem \times Y^\sem} & {Y^\sem} \\
    & {F(X^\syn) \times F(Y^\syn)} \\
    {F(X^\syn)} & {F(X^\syn \times Y^\syn)} & {F(Y^\syn)}
    \arrow[from=1-2, to=2-2]
    \arrow["\cong", from=2-2, to=3-2]
    \arrow["{\pi_1}"', from=1-2, to=1-1]
    \arrow["{\pi_2}", from=1-2, to=1-3]
    \arrow["{F(\pi_2)}"', from=3-2, to=3-3]
    \arrow["{F(\pi_1)}", from=3-2, to=3-1]
    \arrow["{X^\downarrow}"', from=1-1, to=3-1]
    \arrow["{Y^\downarrow}", from=1-3, to=3-3]
\end{tikzcd}\]
The unit type corresponds to the terminal object \(1 \to F(1)\).
Function types are interpreted by the left column of this diagram.
\[\begin{tikzcd}[column sep=0]
    W && {X^\sem \to Y^\sem} \\
    {F(X^\syn \to Y^\syn)} && {X^\sem \to F(Y^\syn)} \\
    & {F(X^\syn) \to F(Y^\syn)}
    \arrow[from=1-3, to=2-3]
    \arrow[dashed, from=2-1, to=3-2]
    \arrow[from=3-2, to=2-3]
    \arrow[from=1-1, to=2-1]
    \arrow[from=1-1, to=1-3]
    \arrow[from=2-1, to=2-3]
    \arrow["\lrcorner"{anchor=center, pos=0.05}, draw=none, from=1-1, to=3-2]
\end{tikzcd}\]
Here, the dashed arrow is the naturally induced one.
Note that since \(F\) does not necessarily preserve exponentials,
it is not an isomorphism.
The universal properties can be verified by direct calculation.
We relied on pullbacks in \(\mathcal F\), which exist in every topos.

As one can see, simple concepts in the internal language quickly gets
complicated when interpreting to the external world.
This is exactly the strength of internal languages.
We only need to care about the interpretations of the premises and the conclusions,
and we can perform all the intermediate steps in a simplistic yet powerful way.

There are other objects in the language such as subobject classifiers,
quotients and inductive types,
but we leave those to standard references such as \textcite{johnstone:2008:elephant}.
These notes will use them but
do not depend on the complicated external constructions of them.
We will also assume infinitely many universes of types,
denoted \(\Omega \of \iType_0 \of \iType_1 \of \cdots\),
where \(\Omega\) is the type of propositions.
If the reader doesn't care, just assume \(\iType \of \iType\).

An object in \(\mathcal G\) consists of three parts
\(X^\sem \xrightarrow{X^\downarrow} F(X^\syn)\).
We naturally would like to access these parts in the internal language.
However, they live in another category,
so it is not immediately clear how to use them.
Luckily, the topos \(\mathcal G\) actually contains copies
of \(\mathcal E\) and \(\mathcal F\).

Consider objects \(X\) in \(\mathcal G\) such that
\(X^\downarrow\) is the identity morphism.
In other words, we have \(X^\sem = F(X^\syn)\).%
\footnote{If you prefer to respect isomorphisms,
then consider \(X\) such that \(X^\downarrow\) is an iso.}
These objects only contain information at the \(\Msyn\) part,
so we call them \(\Msyn\)-modal objects.
They form a full subcategory equivalent to \(\mathcal E\).
Recall that we intend to take \(\mathcal E\) as \(\C{Psh}(\mathcal T)\)
in synthetic Tait computability,
so \(\Msyn\)-modal objects only contain ``syntactic'' information.

Similarly, consider objects \(X\) such that
\(X^\syn = 1\) is the terminal object in \(\mathcal E\).
Since \(F\) is finitely continuous \(F(1)\) is so too,
and \(X^\downarrow\) is unique.
These objects only contain intormation at \(\Msem\),
so analogously we call them \(\Msem\)-modal objects.
They form a full subcategory equivalent to \(\mathcal F\).
These objects contain purely ``semantic'' information.

For every object \(X \of \mathcal G\),
we can replace \(X^\sem\) with \(F(X^\syn)\) to get a new object \(\Msyn X\).
On the other hand, we can replace \(X^\syn\) with \(1\).
This trivializes the syntactic data obtaining \(\Msem X\).
These two operations give us the access that we wanted,
while staying inside \(\mathcal G\).
\[\begin{tikzcd}
{X^\sem} & {X^\sem} & {F(X^\syn)} \\
{F(1)} & {F(X^\syn)} & {F(X^\syn)} \\[-1em]
\textcolor{gray}{\Msem X} & \textcolor{gray}{X} & \textcolor{gray}{\Msyn X}
\arrow["{X^\downarrow}", from=1-2, to=2-2]
\arrow["{\textrm{id}}", from=1-3, to=2-3]
\arrow["{!}", from=1-1, to=2-1]
\arrow[from=1-2, to=1-1]
\arrow["{X^\downarrow}", from=1-2, to=1-3]
\arrow[from=2-2, to=2-3]
\arrow["{F(!)}", from=2-2, to=2-1]
\arrow[color=gray, from=3-2, to=3-1]
\arrow[color=gray, from=3-2, to=3-3]
\end{tikzcd}\]

Internally, we will postulate these two as operators on the types.
But to actually be useful we also have to put in a characterization of them,
otherwise they would just be some random operators with no provable properties.
In fact there is a concise internal way to characterize them.

Consider the object \(\Psyn\) defined by the unique morphism \(0 \to F(1)\).
Here \(0\) is the initial object, which exists in every topos.
This is a subobject of \(1\),
hence it corresponds to a \emph{proposition} in the internal language.
Therefore in the internal language all of its elements will be equal,
and we will use \(\something\) to represent any expression of this type.
\(\Psyn\) stands for ``syntax'', and
being under the assumption \(\Psyn\)
intuitively means that we are only talking about the syntactic part.

Recall that \(U^0 \cong 1\) for any \(U\),
we can calculate the exponential \(\Psyn \to X\) which degenerates to
\[\begin{tikzcd}
\bullet & {0\to X^\sem} & \bullet & 1 \\
{F(1 \to X^\syn)} & {0 \to F(X^\syn)} & {F(X^\syn)} & 1
\arrow[from=2-1, to=2-2]
\arrow[""{name=0, anchor=center, inner sep=0}, from=1-2, to=2-2]
\arrow[from=1-1, to=2-1]
\arrow[from=1-1, to=1-2]
\arrow[from=1-4, to=2-4]
\arrow[from=2-3, to=2-4]
\arrow[""{name=1, anchor=center, inner sep=0}, from=1-3, to=2-3]
\arrow[from=1-3, to=1-4]
\arrow["\lrcorner"{anchor=center, pos=0.125}, draw=none, from=1-1, to=2-2]
\arrow["\lrcorner"{anchor=center, pos=0.125}, draw=none, from=1-3, to=2-4]
\arrow[shorten <=30pt, shorten >=30pt, Rightarrow, from=0, to=1]
\end{tikzcd}\]
Since pulling back along an isomorphism always gives an isomorphism,
we see that \((\Psyn \to X) \cong \Msyn X\).
Thus in the internal language
we can simply postulate an unspecified proposition \(\Psyn\),
and \emph{define} \(\Msyn X = (\Psyn \to X)\).
This simple set-up is enough to produce intricate structures required for our proofs.

On the other hand, \(\Msem X\) can be expressed by the \emph{join} operation
\(\Psyn \star X = (\Psyn +_{\Psyn \times X} X)\)
which is the pushout of
\(\Psyn \leftarrow \Psyn \times X \rightarrow X\).
In the internal language, this is a quotient of \(\Psyn + X\)
by the equivalence relation generated by
\(\ct{inl}(\something) \sim \ct{inr}(x)\).
This can be verified using the fact that limits and colimits
are computed itemwise for \(\mathcal G\).
The join operation (and more generally pushouts)
can be expressed as a quotient inductive type.
We adopt an Agda-like syntax here.
\begin{internal}
\>[0]\kw{inductive} \(\Psyn \star X\) \kw{where}\\
\>[0][@{}l@{\AgdaIndent{0}}]%
\>[4]\(\ct{inl} \of \Psyn \to \Psyn \star X\)\\
\>[4]\(\ct{inr} \of X \to \Psyn \star X\)\\
\>[4]\(\ct{eq} \of \{\Psyn\} (x \of X) \to \ct{inl}(\something) = \ct{inr}(x)\)
\end{internal}

Let's give a test firing of our internal logic.
It should be clear that \(\Msyn \Msem X \cong 1\) by definition in the external world.
Can we prove that in the internal world?
We only need to prove that this type is inhabited,
and that all of its elements are equal.
\(\Msyn \Msem X\) can be written as \(\Psyn \to \Psyn \star X\).
So it is definitely inhabited.
\begin{internal}
\>[0]\(\vb{inhabited} \of \Msyn \Msem X\)\\
\>[0]\(\vb{inhabited} = \ct{inl}\)
\end{internal}
By function extensionality, we only need to prove that
any expression \(u(\something)\) of type \(\Psyn \star X\)
is equal to \(\ct{inl}(\something)\), given a variable \(\something \of \Psyn\).
Performing a case split we have
\begin{itemize}
\item \(u(\something)\) is of the form \(\ct{inl}(\something)\),
then since \(\Psyn\) is a proposition the equality is proved.
\item \(u(\something) = \ct{inr}(x)\),
then by \(\ct{eq}(x)\) we prove the desired equality.
\end{itemize}
This finishes the proof that \(\Msyn \Msem X \cong 1\).
Obviously the underscores are a bit annoying.
So unless for clarification we will suppress all arguments
and $\lambda$ abstractions of type \(\Psyn\).

\begin{comment}
\begin{lemma}[Fracture]
The pullback of
\[\Msem X \longrightarrow \Msem \Msyn X \longleftarrow \Msyn X\]
is again \(X\) itself.
\end{lemma}
\begin{proof}
Let's first clarify what the implied morphisms are.
\(\Msyn X \to \Msem \Msyn X\) is given by the map \(\ct{inr}\) of
the quotient inductive type \(\Psyn \star \Msyn X\);
similarly there is a map \(X \to \Msem X\).
We also have a map \(X \to \Msyn X\):
Recall that \(\Msyn X \cong (\Psyn \to X)\),
therefore given \(x\of X\), we have the constant map
\((\lambda\something\bind x) \of \Psyn \to X\).

\end{proof}
\end{comment}

\begin{idea}
Using the rich internal language of the glued topos,
one can avoid dealing with excessive details of naturality,
which couldn't be handwaved away.
\end{idea}

\subsection{Judgemental, Semantic, and External}

It is time to slow down for an important disambiguation.
There are many concepts that have been assigned the name
``semantic'', ``external'' or ``meta'' in type theory.
While sometimes these words are synonymous,
here we are using these words to refer to different groups of concepts.

First, there are the \emph{judgemental} concepts.
Typing judgements \(x \of A\) and judgemental equalities \(x = y \of A\)
are often described as in the metalanguage,
in contrast to concepts such as \emph{propositional} (or \emph{typal}) equalities.
However these concepts are much better viewed as purely syntactic.
We will more adequately explain the reason in \Cref{sec:objective}.

Second, the models or interpretations of type theory,
in our case the computability structures, are semantic concepts.
The distinction of syntax and semantics is that
the former concerns the emergent properties of the \emph{initial} model.%
\footnote{It is also sometimes said that there is an equivalence between syntax and semantics.
The context of such a statement is, for example, discussing
the equivalence of $\lambda$-theories and Cartesian closed categories.
Note how this is talking about the category of all $\lambda$-theories and
translations between them, instead of a single type theory in our case.}

Third, there are the \emph{external} concepts
(not to be confused with \emph{extensional} concepts)
as opposed to \emph{internal} ones.
Here, our internal language refers to that of the glued topos.
It is different from the language of the original type theory,
in this case simply typed lambda calculus.
The external language is simply informal set theory.
There is an \emph{intended} interpretation of
this internal language into the external world.
As a type theory itself, it can also have different interpretations
just like how simply typed lambda calculus can be interpreted into any model,
but we are not interested in any of those here.

Finally, the gluing introduces the interplay between \(\Msyn\) and \(\Msem\).
It's quite hard to come up with suggestive names without causing name clashes.
\textcite{sterling:2021:thesis} described them as
\emph{separating an object into its syntactic and semantic aspects}.
(There are other uses of gluing apart from proving canonicity and normalization,
and they are given different names in those circumstances.)
In another work,
\textcite{sterling:2022:naive} uses the words ``object'' and ``meta'' instead.

An alternative solution is to borrow from geometry and refer to them
as \emph{open} and \emph{closed} modalities respectively.
We might also call them \emph{downstairs} and \emph{upstairs}
because this is the orientation drawn in most diagrams.
But it would be quite confusing for those who prefer
reading papers in exotic postures.

\subsection{Dance of Modalities}

Now that we've had a first taste of the internal language,
let's give a more systematic treatment of these modalities.
We postulate a proposition in our internal language,
which is interpreted as \(0 \to F(1)\) in the external world.
\begin{internal}
\>[0]\kw{postulate}\\
\>[0][@{}l@{\AgdaIndent{0}}]%
\>[4]\(\Psyn \of \Omega\)
\end{internal}
We said above that we could define \(\Msyn X\) as a function type.
However, for the sake of conceptual clarity,
we prefer to take \(\Msyn X\) as a primitive type, merely isomorphic to the function type.
But in practice, we will implicitly coerce between these two types.

% We postulate that \(\Psyn \to (\Msyn X = X)\), which is saying that
% the \(\Msyn\) part of the type \(X\) is \emph{strictly} equal to that of \(\Msyn X\).
% Remember, having the assumption \(\Psyn\) is equivalent to talking only about the \(\Msyn\) part.
\begin{internal}
\>[0]\kw{postulate}\\
\>[0][@{}l@{\AgdaIndent{0}}]%
\>[4]\({\Msyn} \of \iType_i \to \iType_i\)\\
\>[4]\(\something \of \Msyn X \cong (\Psyn \to X)\)%\\
% \>[4]\(\something \of \Psyn \to (\Msyn X = X)\)
\end{internal}
A type is \(\Msyn\)-modal when the obvious map \(\eta : X \to \Msyn X\) is an isomorphism.
So \(\Msyn X\) is always \(\Msyn\)-modal.
The reader should verify that this means \(X^\downarrow\) is an isomorphism externally.
If this is the case, we denote \(\eta^{-1} : \Msyn X \to X\) the inverse map.

Recall that we can define \(\Msem X\) to be \(\Psyn \star X\).
Again, we take that as an isomorphism instead.
\begin{internal}
\>[0]\kw{postulate}\\
\>[0][@{}l@{\AgdaIndent{0}}]%
\>[4]\({\Msem} \of \iType_i \to \iType_i\)\\
\>[4]\(\something \of \Msem X \cong (\Psyn \star X)\)
\end{internal}
A type \(X\) is defined to be \(\Msem\)-modal when the following equivalent conditions hold:
\begin{itemize}
\item The projection map \(\Psyn \times X \to \Psyn\) is an isomorphism.
\item \(\Msyn X \cong 1\), or \((\Psyn \to X) \cong 1\).
\item \(\Msyn (X \cong 1)\), or \(\Psyn \to (X \cong 1)\).
\item The obvious map \(X \to \Msem X\) is an isomorphism.
\end{itemize}
The proof of equivalence is simple and formalizable.
The reader is encouraged to do it on their own.
Note that we are using an extensional type theory,
which is roughly equivalent to Martin-Löf type theory
with function extensionality and the uniqueness of identity proofs.

In a sense, \(\Msyn\) and \(\Msem\) are complements of each other.
In geometry, this kind of relationship is closely connected to
the behaviour of open and closed subspaces.
Hence in literature they are also called open and closed modalities, respectively.

% (...) Partial types and terms

\subsection{Canonicity}

For canonicity, we take the glued topos
\(\mathcal G = \C{Set} \CommaCat \GSect\),
where \(\GSect \of \C{Psh}(\mathcal T) \to \C{Set}\)
is the closed terms functor \(\GSect(X) = \hom(1, X)\),
or equivalently as \(\yo(1) = 1\), defined by \(\GSect(X) = X(1)\).

We take the set of types \(\Type_{\mathcal G}\) to be
the syntactic types \(\Type_{\mathcal T}\).
This is possible because there is no computation happening on the types.
We now start to assign an object \(\El(A)\) for each \(A \in \Type_{\mathcal G}\).
In the internal language, we need to construct types \(\El_{\mathcal G}(A)\),%
\footnote{Since \(\Type_{\mathcal G}\) is an external set,
this can be viewed as an infinite list of type declarations,
one for each \(A\).
Alternatively one can view this as a macro or template that generates these types.}
while being given \(\Msyn\)-modal types \(\El_{\mathcal T}(A)\).

In addition, our proof strategy by \Cref{lemma:quasiprojectivity}
demands the projection \(\pi : \mathcal G \to \C{Psh}(\mathcal T)\)
to preserve STLC structure.
In other words, the \(\Msyn\)-component of \(\El_{\mathcal G}(A)\)
must be equal to \(\El_{\mathcal T}(A)\).
Synthetically, \(\Psyn \to \El_{\mathcal G}(A) = \El_{\mathcal T}(A)\).

We can express this sort of requirement using a \(\Sigma\) type
\[\sum_{x \of X} \Psyn \to x = y.\]
Let's introduce a notation \(\ext{X}{y}\),
which means ``an element of \(X\) that is equal to \(y\) under \(\Psyn\).''
We will treat this as a subtype of \(X\), with implicit coercion.
So what we want is \(\El_{\mathcal G}(A) \of \ext{\iType}{\El_{\mathcal T}(A)}\).

% Allow partial \(y\)

There is a slight problem though,
namely that most of the time we can only achieve isomorphic types,
not exactly equal types.
Nevertheless, externally we know there is no problem.
Take a partial type \(U\) corresponding to an object in \(\mathcal E\).
The statement that under \(\Psyn\) we have \(X \cong U\)
means \(X^\syn\) and \(U\) are isomorphic in \(\mathcal E\).
Therefore we can replace \(X^\syn\) by \(U\)
to obtain a new object \(X' \of \mathcal G\).%
\footnote{This reasoning is not entirely rigorous.
See \textcite{gratzer:2022:strict} for more information.
They prove a stronger result that applies to arbitrary propositions
instead of just $\Psyn$.}

To codify such a principle, we can introduce a new type.
It is very similar to the \(\textsf{Glue}\) type found in cubical type theories.
If we have a type \(X\) and a partial isomorph \(A\),
we can produce a new type isomorphic to \(X\) and
\emph{equal} to \(A\) under \(\Psyn\).
However this form is not particularly ergonomic for our use case,
and it would produce a lot of boilerplate.

Most of the time,
we have a \(\Msyn\)-modal type \(A\) representing the purely syntactic part,
and a family of \(\Msem\)-modal types \(x\Of A \vdash B(x) \of \iType\).
In this case, the \(\Sigma\) type
\[X = \sum_{x \of A} B(x)\]
will satisfy \(\Psyn \to (X \cong A)\).
The reader can try to prove this, perhaps in a proof assistant.
We should have a new type \(X'\) isomorphic to \(X\) and
\(\Psyn \to (X' = A)\).
Our notation will be optimized for this particular use case.

We postulate an alternative version of \(\Sigma\) type.
Following \textcite{sterling:2023:geometry}, we call it the G type.
It is has similar rules to \(\Sigma\) types,
except we add appropriate equations when under \(\Psyn\).
We write out part of them in \Cref{fig:Gtype}.
There's a nice parallel in our notations for big operators.
\begin{center}
\begin{tabular}{c|c}
Nuprl-style & Big operator \\ \hline
\((x \of A) \to B(x)\) & \(\prod_{x\of A} B(x)\) \\[2pt]
\((x \of A) \times B(x)\) & \(\sum_{x\of A} B(x)\) \\[2pt]
\((x \of A) \ltimes B(x)\) & \(\Refine_{x\of A} B(x)\) \\
\end{tabular}
\end{center}
And in fact we have
\[\Refine_{x\of A} B(x) \cong \sum_{x\of A} B(x).\]
Given \(p \of \Refine_{x \of A} B(x)\), we refer to the syntactic component as
\(p^\syn \of A\), and the semantic component as \(p^\sem \of B(p^\syn)\).

\begin{figure}
\begin{mathpar}
\inferrule{A \of \iType
\and \text{\(A\) is \(\Msyn\)-modal}
\and x \Of A \vdash B \of \iType
\and x \Of A \vdash \text{\(B\) is \(\Msem\)-modal}}
{\Refine_{x \of A} B \of \iType} \and
\inferrule{\text{(same assumptions) \and \text{\(\Psyn\) holds}}}
{A = \Refine_{x \of A} B} \and
\inferrule{a \of A
\and b \of B[x/a]}
{[a, b] \of \Refine_{x \of A} B} \and
\inferrule{a \of A
\and b \of B[x/a]
\and \text{\(\Psyn\) holds}}
{a = [a, b] \of A} \and
\text{(Projection and \(\beta, \eta\) omitted)}
\end{mathpar}
\caption{Rules of the G type}
\label{fig:Gtype}
\end{figure}

\begin{theorem}[Canonicity]
For \(\vdash t \of \Ans\), either \(\vdash t = \yes\of\Ans\)
or \(\vdash t = \no\of\Ans\).
\end{theorem}
\begin{proof}
Let's review what our goals are.
We defined \(\mathcal G = \textrm{Id}_{\C{Set}} \CommaCat \C{Psh}(\mathcal T)\).
And we wish to construct a STLC structure on it, so that
we can use \Cref{lemma:quasiprojectivity}.
We already have \(\Msyn\)-modal types \(\El_{\mathcal T}(A)\),
and we need to assign types
\[\El_{\mathcal G}(A) \of \ext{\iType}{\El_{\mathcal T}(A)}.\]
We also need elements such as \(\synct{yes}_{\mathcal G}, \synct{lam}_{\mathcal G}\)
that restricts to the corresponding stuff in \(\El_{\mathcal T}\) under \(\Psyn\).
The development in internal language proceeds like writing code in Agda.
For the product type we define
\begin{internal}
\>[0]\(\El_{\mathcal G}(A \times B) \of \ext{\iType}{\El_{\mathcal T}(A \times B)}\)\\
\>[0]\(\El_{\mathcal G}(A \times B) = \Refine_{p \of \El_{\mathcal T}(A \times B)} \hole\)
\end{internal}
The hole should be filled with a type that is \(\Msem\)-modal.
Recall that the G type is isomorphic to the \(\Sigma\) type,
and we need to have \(\El_{\mathcal G}(A \times B) \cong \El_{\mathcal G}(A) \times \El_{\mathcal G}(B)\),
so we fill in
\begin{internal}
\>[0]\(\El_{\mathcal G}(A \times B) \of \ext{\iType}{\El_{\mathcal T}(A \times B)}\)\\
\>[0]\(\El_{\mathcal G}(A \times B) =\)\\
\>[0][@{}l@{\AgdaIndent{0}}]%
\>[4]\(\Refine_{p \of \El_{\mathcal T}(A \times B)}
\sum_{\tilde p \of \El_{\mathcal G}(A) \times \El_{\mathcal G}(B)}
\Psyn \to \tilde p = (\synct{fst}_{\mathcal T}(p), \synct{snd}_{\mathcal T}(p))\)
\end{internal}
This is equivalently written as the product of
\(\ext{\El_{\mathcal G}(A)}{\synct{fst}(p)}\) and \(\ext{\El_{\mathcal G}(B)}{\synct{snd}(p)}\).
This is indeed \(\Msem\)-modal.
\begin{internal}
\>[0]\(\synct{pair}_{\mathcal G}(a, b) = [\synct{pair}_{\mathcal T}(a^\syn, b^\syn),
((a, b), \lambda \something\bind \ct{refl})]\)\\
\>[0]\(\synct{fst}_{\mathcal G}(p) = \pi_1(\pi_1(p^\sem))\)\\
\>[0]\(\synct{snd}_{\mathcal G}(p) = \pi_2(\pi_1(p^\sem))\)
\end{internal}

For function types we similarly fill in
\begin{internal}
\>[0]\(\El_{\mathcal G}(A \to B) \of \ext{\iType}{\El_{\mathcal T}(A \to B)}\)\\
\>[0]\(\El_{\mathcal G}(A \to B) =\)\\
\>[0][@{}l@{\AgdaIndent{0}}]%
\>[4]\(\Refine_{f \of \El_{\mathcal T}(A \to B)}
\sum_{\tilde f \of \El_{\mathcal G}(A) \to \El_{\mathcal G}(B)}
\Psyn \to \tilde f = \lambda x \bind \synct{app}_{\mathcal T}(f, x)\)
\end{internal}
The variable \(x\) has type \(\El_{\mathcal G}(A)\),
which is equal to \(\El_{\mathcal T}(A)\) under \(\Psyn\),
so everything checks out.
\begin{internal}
\>[0]\(\synct{lam}_{\mathcal G}(f) =
[\synct{lam}_{\mathcal T}(\lambda x \bind \hole), (f, \lambda\something\bind \hole)]\)
\end{internal}
We are given \(f \of \El_{\mathcal G}(A) \to \El_{\mathcal G}(B)\) and
\(x \of \El_{\mathcal T}(A)\).
The first hole we need to fill in is of type \(\El_{\mathcal T}(B)\).
Since the type is \(\Msyn\)-modal, it is isomorphic to
\(\Psyn \to \El_{\mathcal T}(B)\).
Now that we are given an assumption \(\Psyn\),
we have \(\El_{\mathcal G} = \El_{\mathcal T}\).
Therefore \(f(x)\) typechecks and fits the bill.
\begin{internal}
\>[0]\(\synct{lam}_{\mathcal G}(f) =
[\synct{lam}_{\mathcal T}(\lambda x \bind \eta^{-1}(f(x))), (f, \lambda\something\bind \hole)]\)
\end{internal}
The second hole comes from equational reasoning using the syntactic \(\beta\) law.
We leave as an exercise to construct \(\synct{app}\) and prove the equations.

Note that these are rigorously all we need, no more handwaving of naturality conditions!
In addition, these kinds of logic are very suitable for proof assistants to check.
It has already been \href{https://github.com/jonsterling/agda-stc}{hacked into Agda}.

We leave the unit type as an exercise.
For the \(\Ans\) type, we need
\begin{internal}
\>[0]\(\El_{\mathcal G}(\Ans) \of \ext{\iType}{\El_{\mathcal T}(\Ans)}\)\\
\>[0]\(\El_{\mathcal G}(\Ans) = \Refine_{p \of \El_{\mathcal T}(\Ans)}
    \Msem ((p = \synct{yes}_{\mathcal T}) + (p = \synct{no}_{\mathcal T}))\)
\end{internal}
This is the interesting part.
Remember that \(\Msem X\) is always \(\Msem\)-modal,
which ensures that we can use the G type.
\begin{internal}
\>[0]\(\synct{yes}_{\mathcal G} \of \ext{\El_{\mathcal G}(\Ans)}{\synct{yes}_{\mathcal T}}\)\\
\>[0]\(\synct{yes}_{\mathcal G} = [\synct{yes}_{\mathcal T}, \ct{inr}(\ct{inl}(\ct{refl}))]\)\\
% \>[0]\\
\>[0]\(\synct{no}_{\mathcal G} \of \ext{\El_{\mathcal G}(\Ans)}{\synct{no}_{\mathcal T}}\)\\
\>[0]\(\synct{no}_{\mathcal G} = [\synct{no}_{\mathcal T}, \ct{inr}(\ct{inr}(\ct{refl}))]\)
\end{internal}
And now it's done! This constructs a STLC structure over \(\mathcal G\),
so we can conclude that there is a map \(\Sem{-}\)
that will interpret any term \(\vdash t \of \Ans\)
as a morphism \(\hom_{\mathcal G}(1, \El_{\mathcal G}(\Ans))\).
In the internal language, this is a closed element of \(\El_{\mathcal G}(\Ans)\),
which would have two components \([t, u]\),
where we know the first component is equal to \(t\).
\(u\) must therefore be an element of the type
\(\Msem ((t = \synct{yes}_{\mathcal T}) + (t = \synct{no}_{\mathcal T}))\).
Let's break it down by parts.

As a proposition, if \(t = \yes\) is true,%
\footnote{We use \(\yes\) in the external language,
and \(\synct{yes}_{\mathcal T}\) in the internal language.}
it would correspond to the object \(1 \of \mathcal G\),
and otherwise it would correspond to \(0 \of \mathcal G\).
Taking the coproduct gives us the object
\[\begin{tikzcd}
{\{\atom{t} \mid t = \yes\} \cup \{\atom{f} \mid t = \no\}} \\
{F(\text{\dots})}
\arrow[from=1-1, to=2-1]
\end{tikzcd}\]
where we use \(\atom{t}, \atom{f}\) to label the branches of the coproduct in \(\C{Set}\).
We don't care about what's downstairs because we immediately wipe it out
with \(\Msem\), which produces
\[\begin{tikzcd}
{\{\atom{t} \mid t = \yes\} \cup \{\atom{f} \mid t = \no\}} \\
{F(1)}
\arrow[from=1-1, to=2-1]
\end{tikzcd}\]
A closed element of this object would be
\[\begin{tikzcd}
{1} & {\{\atom{t} \mid t = \yes\} \cup \{\atom{f} \mid t = \no\}} \\
{F(1)} & {F(1)}
\arrow[from=1-1, to=2-1]
\arrow[from=1-2, to=2-2]
\arrow[from=1-1, to=1-2]
\arrow[from=2-1, to=2-2]
\end{tikzcd}\]
and the morphism upstairs completes the proof.
\end{proof}
\begin{remark}
After becoming more accustomed to the internal language of gluing,
the choice of \(\El_{\mathcal G}(\Ans)\) will become obvious.
The reader is encouraged to not think about the expression
in relation to its externalization,
but rather to think purely internally.
\end{remark}

\subsection{Normalization}

Recall \(\mathcal A\) is the category of contexts and renamings,
and \(\rho : \mathcal A \hookrightarrow \mathcal T\) is the inclusion.
Previously we used the functor \(\Nerve(\Gamma) = \hom(\rho(-), \Gamma)\),
which is a functor \(\mathcal T \to \C{Psh}(\mathcal A)\).
Now we need to upgrade to \(\C{Psh}(\mathcal T)\) too.
The natural generalization is then
\[\rho^*(X) = X \comp \rho.\]
\begin{remark}
This is the \textbf{Yoneda extension} of the functor \(\Nerve\),
because it is the unique cocontinuous functor making this diagram commute.
\[\begin{tikzcd}
{\C{Psh}(\mathcal A)} \\
{\mathcal T} & {\C{Psh}(\mathcal T)}
\arrow["\Nerve", from=2-1, to=1-1]
\arrow["\yo"', from=2-1, to=2-2]
\arrow[dashed, from=2-2, to=1-1]
\end{tikzcd}\]
This property is why we call \(\C{Psh}\) the \emph{free cocompletion}.
\end{remark}

As expected, we take
\(\mathcal G = \textrm{Id}_{\C{Psh}(\mathcal A)}\CommaCat \rho^*\)
to be our glued topos for normalization.
Its objects are of the form
\[X^\sem \xrightarrow{\smash{\;X^\downarrow\;}} \rho^*(X^\syn)\]
with \(X^\sem\) a presheaf over \(\mathcal A\),
and \(X^\syn\) a presheaf over \(\mathcal T\).

We need to be able to reason about normal and neutral forms
in the internal language.
We know that normal forms can't be a presheaf over \(\mathcal T\),
because they are not stable under substitution.
But they are stable under renamings.
And so we can construct an object \(\Elnf(A)\) in \(\mathcal G\),
with the \(\Msyn\) part equal to the presheaf of terms over \(\mathcal T\),
and the \(\Msem\) part equal to the presheaf of normal forms over \(\mathcal A\).
The morphism is the obvious inclusion of normal forms into terms.
Similarly we have \(\Elvar(A)\) and \(\Elne(A)\) too.
There is of course a map \(\nect{var} \of \Elvar(A) \to \Elne(A)\).
We will mention other operations on normal and neutral forms on our way.

This is another manifestation of the power of gluing.
It allows us to talk about concepts such as normal forms,
which don't make sense in the syntactic world
because we quotiented by judgemental equality.
In addition, the language automatically tracks
the correspondence between normal forms and terms,
so we don't need to worry about the normalization function
producing a normal form for the wrong term!
Previously this was proved by a separate round of induction over the whole thing.

\begin{idea}
Inside the synthetic language,
the normal and neutral forms are realized as computability spaces over the terms.
Their connection with the syntax becomes integrated into the structure of the proof,
instead of an afterthought.
\end{idea}

Recall that apart from constructing a model,
we also need to define reification and reflection.
We can do all of these \emph{à la carte}.
\begin{internal}
\>[0]\(\reflect^A \of \ext{\Elne(A) \to \El_{\mathcal G}(A)}{\textrm{id}}\) \\
\>[0]\(\reify^A \of \ext{\El_{\mathcal G}(A) \to \Elnf(A)}{\textrm{id}}\)
\end{internal}
You should convince yourself that this definition makes sense.
In the external language, this unfolds to
a reflection function that maps neutral forms to computability structures,
and a reification function that maps computability structures to normal forms.
Requiring the syntactic part of them to be \(\textrm{id}\) ensures
the a neutral form corresponding to the term \(t\) is mapped to
a computability structure also corresponding to \(t\), etc.

\begin{theorem}[Normalization]
There is an isomorphism
\[\normalize \of \El_{\mathcal T}(A) \cong \Elnf(A)\]
such that under \(\Psyn\) we have \(\normalize = \textrm{id}\).
\end{theorem}
\begin{proof}
The syntactic definition of neutral and normal forms
gives rise to internal maps that operate on \(\Elnf\) and \(\Elne\).
For instance, for product types we are given maps
\begin{internal}
\>[0]\(\nfct{pair} \of \Elnf(A) \times \Elnf(B) \to \Elnf(A \times B)\)\\
\>[0]\(\nect{fst} \of \Elne(A \times B) \to \Elne(A)\)\\
\>[0]\(\nect{snd} \of \Elne(A \times B) \to \Elne(B)\)
\end{internal}
such that under \(\Psyn\), \(\nfct{pair}\) restricts to the syntactic pairing function
\(\synct{pair}_{\mathcal T}\), etc.
The gluing for \(\El_{\mathcal G}\) is much the same as in canonicity, and we define
(eliding parentheses after \(\reflect\) or \(\reify\),
so they are parsed like negation \(\neg\))
\begin{internal}
\>[0]\(\reflect^{A \times B}(p) =
    \synct{pair}_{\mathcal G}(\reflect^A \nect{fst}(p), \reflect^B\nect{snd}(p))\)\\
\>[0]\(\reify^{A \times B}(p) =
    \nfct{pair}(\reify^A \synct{fst}_{\mathcal G}(p), \reify^B\synct{snd}_{\mathcal G}(p))\)
\end{internal}
The \(\Unit\) type is left as an exercise, and for functions we are given
\begin{internal}
\>[0]\(\nfct{lam} \of (\Elvar(A) \to \Elnf(B)) \to \Elnf(A \to B)\)\\
\>[0]\(\nect{app} \of \Elne(A \to B) \times \Elnf(A) \to \Elne(B)\)
\end{internal}
Observe that $\nfct{lam}$ takes a \(\Elvar\) here.
The \emph{analytic} justification is that its interpretation
in the external language is the correct one.
It can also be \emph{synthetically} justified since
$\lambda$ bindings in normal forms are only supposed to be
substituted with variables, not terms,
otherwise it would produce a non-normal term.
\begin{internal}
\>[0]\(\reflect^{A \to B}(f) =
    \synct{lam}_{\mathcal G}(\lambda x\bind \reflect^B \nect{app}(f, \reify^A x))\)\\
\>[0]\(\reify^{A \to B}(f) =
    \nfct{lam}(\lambda x\bind \reify^B \synct{app}_{\mathcal G}(f, \reflect^A \nect{var}(x)))\)
\end{internal}
Our naïve proof is much more sloppy and carries a lot of renaming annotations around,
and the categorical proof either handwaves the naturality conditions away
or is very lengthy and complicated.
Synthetic Tait computability handles this elegantly,
including the problem of binding.
\begin{internal}
\>[0]\(\nfct{yes}, \nfct{no} \of \Elnf(\Ans)\)\\
\>[0]\(\nect{up} \of \Elne(\Ans) \to \Elnf(\Ans)\)
\end{internal}
Remember that \(\Ans\) has the rule that neutral forms of this type are also neutral,
and other types don't have this rule to force \(\eta\) expansion.
We make this as an explicit conversion operator here.
Under \(\Psyn\), \(\nect{up}\) will restrict to the identity function
\(\El_{\mathcal T}(\Ans) \to \El_{\mathcal T}(\Ans)\).

The previous constructions for \(\El_{\mathcal G}\)
are all identical to those in the canonicity proof,
so we omitted them.
We define the gluing for \(\Ans\) differently.
Since in the naïve proof we defined a computability structure on \(\Ans\)
as just a normal form of type \(\Ans\),
the only reasonable solution here is
\begin{internal}
\>[0]\(\El_{\mathcal G}(\Ans) = \Elnf(\Ans)\)\\
\>[0]\(\synct{yes}_{\mathcal G} = \nfct{yes}\)\\
\>[0]\(\synct{no}_{\mathcal G} = \nfct{no}\)\\
\>[0]\(\reflect^\Ans(a) = a\)\\
\>[0]\(\reify^\Ans(a) = \nect{up}(a)\)
\end{internal}
This completes the construction of a STLC structure in \(\mathcal G\),
and morphisms \(\reflect\) and \(\reify\).
Now we can reap the fruits.

The interpretation \(\Sem{-}\) maps a term \(\Gamma \vdash t \of A\)
to a morphism \(\Sem{\Gamma} \to \El_{\mathcal G}(A)\).
We know that
\(\Sem{\Gamma}^\syn = \yo(\Gamma)\) by \Cref{lemma:quasiprojectivity},
where \(\yo\) is the Yoneda embedding of \(\mathcal T\).
We also know from \Cref{lemma:quasiprojectivity} that
\(\Sem{\Gamma \cdot A} \cong \Sem{\Gamma} \times \El_{\mathcal G}(A)\),
since \(\Type_{\mathcal G} = \Type_{\mathcal T}\) in our case.
This allows us to compute \(\Sem{\Gamma}\)
because every syntactic context
can be written as \(1 \cdot A_1 \cdot A_2 \cdots A_n\),
and using the isomorphism we see
\[\Sem{\Gamma} \cong \El_{\mathcal G}(A_1) \times \cdots \times \El_{\mathcal G}(A_n).\]

We can construct another computability structure over \(\yo(\Gamma)\).
Consider a morphism
\[\yo_{\mathcal A}(\Gamma) \longrightarrow \rho^*(\yo_{\mathcal T}(\Gamma))\]
The presheaf on the left \(\hom_{\mathcal A}(-, \Gamma)\) consists of the renamings into \(\Gamma\),
whereas the presheaf on the right \(\hom_{\mathcal T}(\rho(-), \Gamma)\) consists of the substitutions into \(\Gamma\).
Therefore we have an obvious embedding.
We denote this structure as \(\Ren{\Gamma}\).
This extends to a functor \(\Ren{-} \of \mathcal A \to \mathcal G\).
\(\Ren{-}\) is to \(\Sem{-}\) as \(\Elvar\) is to \(\El_{\mathcal G}\).
More precisely,
\[\Ren{-} \cong \Elvar(A_1) \times \cdots \times \Elvar(A_n).\]
Therefore we can apply the map
\({\reflect^A} \comp \synct{var} \of \Elvar(A) \to \El_{\mathcal G}(A)\)
componentwise, obtaining a map
\(\reflect^\Gamma \of \Ren{\Gamma} \to \Sem{\Gamma}\).

We have a composite map
\[\begin{tikzcd}
{\Ren{\Gamma}} & {\Sem{\Gamma}} & {\El_{\mathcal G}(A)} & {\Elnf(A)} \\
{\yo(\Gamma)} & {\yo(\Gamma)} & {\El_{\mathcal T}(A)} & {\El_{\mathcal T}(A)}
\arrow["{\reflect^\Gamma}", from=1-1, to=1-2]
\arrow["{\Sem{t}}", from=1-2, to=1-3]
\arrow["{\reify^A}", from=1-3, to=1-4]
\arrow["{\textrm{id}}"', from=2-3, to=2-4]
\arrow["{\textrm{id}}"', from=2-1, to=2-2]
\arrow["t"', from=2-2, to=2-3]
\arrow[maps to, from=1-1, to=2-1]
\arrow[maps to, from=1-2, to=2-2]
\arrow[maps to, from=1-3, to=2-3]
\arrow[maps to, from=1-4, to=2-4]
\end{tikzcd}\]
where the lower row is the \(\Msyn\)-component.
The \(\Msem\)-component gives a natural transformation of presheaves over \(\mathcal A\)
\[\hom_{\mathcal T}(\rho(-), \Gamma) \longrightarrow \Elnf(-, A).\]
Therefore it maps \(\textrm{id} \of \Gamma \to \Gamma\) to a normal form \(\Elnf(\Gamma, A)\).
The rest can be done by diagram chasing.
\end{proof}

\subsection{Peripheral Improvements}

There are some improvements here and there we can make.
Most importantly, since \(\Type\) is a plain set instead of an object in \(\mathcal G\),
it is somewhat cumbersome to deal with it in the internal language.
We didn't do anything about it because in dependent types
\(\Type\) will naturally become another object in \(\mathcal G\),
so the problem solves itself.
% In the dependent case, synthetic Tait computability becomes even more elegant.
% Of course we could have treated simple types as a degenerate case of type dependency.

At the end of the day, we exit the synthetic world
and finish the proof with some external reasoning.
But since internal languages is capable of doing any constructive math,
we can stay internal while doing that,
introducing an \emph{adequacy} axiom that lets us
connect the inductive definition of syntax to the synthetic language.
This is done by \textcite{sterling:2022:naive}.

The finishing bits of the normalization proof can be made more principled
using some language of geometric morphisms between topoi.
For example, we could have simply computed that
\[\hom_{\mathcal G}(\Ren{\Gamma}, X) \cong
\hom_{\C{Psh}(\mathcal A)}(\yo_{\mathcal A}(\Gamma), X^\sem)\]
via the properties of adjunction.
Additionally,
the map \(\reflect^\Gamma \of \Ren{\Gamma} \to \Sem{\Gamma}\)
is constructed by hand,
but the nature of this map can be described using a universal property,
discussed e.g. by \textcite{sterling:2023:geometry},
where it is called the ``hydration'' map.

\section{Dependent Types, Naïvely}

We use Tarski-style universes.
The elements of \(\Univ\) are representative codes for the types
containing \(\Ucode{Ans}, \Ucode{Pi}\) etc.
There is a type former \(\UEl\) that turns codes into types,
with judgemental equalities such as \[\Ans = \UEl(\Ucode{Ans}) \isType\]
connecting the types with the type codes.
This better reveals the structure than Russell-style universes.

\begin{theorem}[Equational consistency]
\(\nvdash \yes = \no \of \Ans\).
\end{theorem}
\begin{proof}
Since types, contexts and terms are now entangled, we need to simultaneously deal with them all.
We assign a set \(\Sem{\Gamma}\) to each context \(\Gamma\),
a family of sets \(\Sem{A}\) indexed by \(\Sem{\Gamma}\) to each type \(\Gamma \vdash A \isType\),
a family of elements \(\Sem{a}\) of \(\Sem{A}\) to each term \(\Gamma \vdash a \of A\),
such that judgementally equal syntax are given equal interpretations.

The empty context is assigned a singleton set.
Context extension is given by disjoint union
\[\Sem{\Gamma \cdot A} = \coprod_{\gamma \of \Sem{\Gamma}} \Sem{A}(\gamma). \]
Compare this with the non-dependent case where \(\Sem{\Gamma \cdot A} = \Sem{\Gamma} \times \Sem{A}\).
The interpretations to types are obvious.
We use disjoint unions of sets to interpret the \(\Sigma\) type,
and similarly dependent function sets for the \(\Pi\) type.
% For \(\Gamma \vdash t = s \isType\), we give
% \[\Sem{t = s}(\gamma) = \begin{cases}
% \{\atom{refl}\} & (\Sem{t}(\gamma) = \Sem{t}(\Gamma))\\
% \varnothing & (\Sem{t}(\gamma) \ne \Sem{t}(\Gamma))
% \end{cases}\]

The universe type requires slightly more care.
What we need is a set of sets that is closed under \(\coprod\) and \(\prod\) operations.
We can draw from the arsenal of set theory,
where \textrm{Grothendieck universes} satisfy our requirements.%
\footnote{This can also be stated purely categorically, and it generalizes to topoi.}
We simply take as an axiom that there are enough Grothendieck universes in our ambient set theory.
Take one of them denoted \(U\) and assign \(\Sem{\Univ}(\gamma) = U\).
Therefore the interpretation of \(c \of \Univ\) is a set inside the universe \(U\),
and we can let \(\Sem{\UEl(c)}(\gamma) = \Sem{c}(\gamma)\).

With this, the argument from the non-dependent case
can be easily adopted here, and thus we established that \(\yes \ne \no\) in the empty context.
\end{proof}
\begin{remark}
If we tried to prove \(\Gamma \nvdash \yes = \no \of \Ans\),
then there is a genuine concern that \(\Sem{\Gamma}\) may be empty.
Normalization would be needed in this case.
% A traditional way to prove it is to define a rewriting system on the terms
% such that judgemental equality is the smallest equivalence relation containing it,
% and prove that it is \emph{confluent},
% i.e. if a term rewrites to two different terms,
% then there is always a way to continue the two paths so that they join again.
% Then, since \(\yes\) and \(\no\) are irreducible,
% they cannot be connected by a zigzag of rewrites,
% and thus cannot be equal judgementally.
\end{remark}

\subsection{Canonicity}

Canonicity scales with no problem to dependent types,
except all the steps are now entangled in a big mutual induction.

\begin{theorem}[Canonicity]\label{theorem:canonicity:dependent}
For every term \(\vdash t \of \Ans\),
either \(\vdash t = \yes \of \Ans\) or \(\vdash t = \no \of \Ans\) is derivable.
\end{theorem}
\begin{proof}
We define a set of computability structures for each closed term.
A computability structure for \(f \of \prod_{x \of A} B\)
maps each closed term \(a \of A\) with computability structure \(u\)
to a computability structure for \(f(a) \of B[x/a]\).
The reader should be able to handle the most cases.
The interesting case is about universes.
We define a computability structure for \(A \of \Univ\)
to be a set \(S_A\) with a map \(\pi_A \of S_A \to \GSect(\UEl(A))\),
i.e. a way to assign computability structures to closed terms.
And a computability structure for \(t \of \UEl(A)\)
can be simply taken as a \(s \in S_A\) such that \(\pi_A(s) = t\).

Now we assign a computability structure recursively for each term.
The non-straightforward case here is just the universe codes.
We define \(\Sem{\Ucode{Ans}}\) to be the assignment of computability structures for \(\Ans\),
since this is required by \(\UEl(\Ucode{Ans}) = \Ans\).
The rest can be similarly defined.

The argument using computability morphisms remain largely the same.
\end{proof}

\subsection{Normalization}

Since dependent types can now involve computation in the types,
the only reasonable move for normalization is to also include
normal and neutral forms for types, written as
\(A \isNfType\) and \(A \isNeType\).
Also, the types of normal or neutral forms need not be normal,
because we need to deal with substitutions etc.
We list the rules for \(\Sigma, \Pi\) and a universe type
in \Cref{fig:dep-nf-ne}.
Here we use \((x \of A) \to B\) and \((x \of A) \times B\) to save space.
\begin{figure}
\begin{mathpar}
\inferrule{ }{\Gamma \vdash \Ans \isNfType} \and
\inferrule{ }{\Gamma \vdash \yes \isNf \Ans} \and
\inferrule{ }{\Gamma \vdash \no \isNf \Ans} \and
\inferrule{\Gamma \vdash t \isNe \Ans}{\Gamma \vdash t \isNf \Ans} \and
\inferrule{\Gamma \vdash A \isNfType \and \Gamma, x\Of A \vdash B \isNfType}
    {\Gamma \vdash (x \of A) \times B \isNfType} \and
\text{(Same for \(\Pi\))} \and
\inferrule{\Gamma \vdash a \isNf A \and \Gamma \vdash b \isNf B[x/a]}
    {\Gamma \vdash (a, b) \isNf (x \of A) \times B} \and
\inferrule{\Gamma \vdash p \isNe (x \of A) \times B}
    {\Gamma \vdash \pi_1(p) \isNe A} \and
\inferrule{\Gamma \vdash p \isNe (x \of A) \times B}
    {\Gamma \vdash \pi_2(p) \isNe B[x/\pi_1(p)]} \and
\inferrule{\Gamma, x \Of A \vdash t \isNf B}
    {\Gamma \vdash \lambda x \bind t \isNf (x \of A) \to B} \and
\inferrule{\Gamma \vdash f \isNe (x \of A) \to B \and
    \Gamma \vdash a \isNf A}
    {\Gamma \vdash f(a) \isNe B} \and
\inferrule{ }{\Gamma \vdash \Univ \isNfType} \and
\inferrule{\Gamma \vdash u \isNe \Univ}{\Gamma \vdash u \isNf\Univ} \and
\inferrule{ }{\Gamma \vdash \Ucode{Ans} \isNf \Univ} \and
\text{(Similar codes omitted)} \and
\inferrule{\Gamma \vdash u \isNe \Univ}{\Gamma \vdash \UEl(u) \isNeType} \and
\inferrule{\Gamma \vdash x \isVar A}{\Gamma \vdash x \isNe A} \and
\inferrule{\Gamma \vdash A \isNeType}{\Gamma \vdash A \isNfType} \and
\inferrule{\Gamma \vdash A \isNeType \and \Gamma \vdash a \isNe A}
    {\Gamma \vdash a \isNf A}
\end{mathpar}
\caption{Dependent normal and neutral forms}
\label{fig:dep-nf-ne}
\end{figure}

Apart from assigning computability structures,
we also assign a normal type to each type.
For example, the normal type corresponding to \(\prod_{x \of A} B\)
is \(\prod_{x \of A'} B'\) where \(A'\) and \(B'\) are the normal types
assigned to \(A\) and \(B\).

The only remarkable case is for the universe.
We define a computability structure for \(\Gamma \vdash A \of \Univ\)
as a normal form equal to \(A\),
together with a way to assign computability structures
to open terms \(\Delta \vdash t \of A[\sigma]\),
where \(\sigma\) is any substitution from \(\Delta\) to \(\Gamma\).

This proof is very tedious, we refer the reader to \textcite{coquand:2018:canonicity}
for the detailed proof.

\section{Objective Syntax}\label{sec:objective}

\emph{Subjective syntax} typically involves using strings of characters or syntax trees etc.
to build the so-called \emph{raw terms} of type theory.
Then substitution, various typing and reduction relations are defined.

\emph{Objective syntax} is the \emph{presentation-invariant} structures of syntax.
The situation is similar to the study of linear algebra without picking a basis.~\cite[\S{} 0.6]{sterling:2021:thesis}
Objective metatheory hides the implementation details,
while exposing the critical features that allows us to establish important theorems.

In this section, we review the objective syntax of dependent type theories.
All of these can, of course, be specialized to the non-dependent case.
We chose not to do that to give a more steady introduction.
% However, the reader can re-read the previous sections with these new ideas
% to get a full

\subsection{Categories with Families}

In dependent types, \(\Type\) is no longer a set independent of the context.
Therefore it will also be a presheaf, just like the terms.
Let \(\Term\) be the presheaf of terms up to \(\beta\eta\) equivalence.
(They are typed, but terms of different types are mixed in the same sets.)
We should have a projection natural transformation
\(\typeof : \Term \to \Type\) that assigns each term its corresponding type.

\begin{remark}
If the reader is uncomfortable with putting terms of different types together,
there is another formalism that requires slightly more category theory.
\(\Term\) can be considered as a presheaf over
the category of types \(\int \Type\).
The two formalisms are equivalent.
\end{remark}

A type \(\Gamma \vdash A \isType\) corresponds to an element
\(A \in \Type(\Gamma)\), which by Yoneda is equivalent to a morphism
\(A \of \yo(\Gamma) \to \Type\).
We also need to deal with context extensions.
In the syntactic category, we have
\[\begin{split}
\hom(\Delta, \Gamma \cdot A) &=
\left\{ (\sigma, a)
\,\middle|\,
\begin{gathered}
\Delta \vdash \sigma \of \Gamma \\
\Delta \vdash a \of A[\sigma]
\end{gathered}
\right\}\\
&=
\left\{ (\sigma, a)
\,\middle|\,
\begin{gathered}
\sigma \of \hom(\Delta, \Gamma) \\
a \of \Term(\Delta) \\
\typeof(a) = A[\sigma]
\end{gathered}
\right\}
\end{split}\]
This suggests expressing \(\yo(\Gamma \cdot A)\) as a pullback
\[\begin{tikzcd}
{\yo(\Gamma \cdot A)} & \Term \\
{\yo(\Gamma)} & \Type
\arrow["\typeof", from=1-2, to=2-2]
\arrow["A"', from=2-1, to=2-2]
\arrow["\mathfrak{q}", from=1-1, to=1-2]
\arrow["\yo(\mathfrak{p})"', from=1-1, to=2-1]
\arrow["\lrcorner"{anchor=center, pos=0.125}, draw=none, from=1-1, to=2-2]
\end{tikzcd}\]
where \(\mathfrak{p}\) is the substitution that forgets one variable,
and \(\mathfrak{q}\) is exactly the variable \(x \of A\) considered as a term.
This situation in the syntactic category inspires us
to define models in an analogous way.
\begin{definition}
A \textbf{category with families} is a category \(\mathcal C\)
equipped with two presheaves and a natural transformation
\(\typeof \of \Term \to \Type\),
such that for every morphism \(A \of \hom(\yo(\Gamma), \Type)\),
its pullback with \(\typeof\) is representable.
\end{definition}
\begin{remark}
More accurately, we require a binary operation \(\Gamma \cdot A\)
that gives the corresponding representing objects.
This is the same issue with the difference of existence of products and
a given binary operation that produces the products.
\end{remark}

As an example, we demonstrate how to add \(\Pi\) types to CwFs.

The type formation rule corresponds to a morphism
\(\Pi \of P \to \Type\), where \(P\) is the presheaf
of ordered pairs
\[P(\Gamma) = \left\{(A, B)\,\middle|\,
\begin{gathered}
A \in \Type(\Gamma) \\
B \in \Type(\Gamma \cdot A)
\end{gathered}
\right\},\]
which records the domain of the type former.
More categorically, we can express this as the exponential object
in the slice category \(\C{Psh}(\mathcal C) / \Type\) of the following two objects:
\[\begin{tikzcd}
\Type\times\Type && \Term \\
& { \Type}
\arrow["{\pi_2}"', from=1-1, to=2-2]
\arrow["\typeof", from=1-3, to=2-2]
\end{tikzcd}\]
The reader can try to calculate and prove that this indeed produces \(P\).
Such a method has the advantage that substitution is automatically taken care of,
whereas defining the sets \(P(\Gamma)\) by hand involves filling in the functoriality details.
Even better, we can express this in the internal language of presheaf topos as
\[P = \sum_{A \of \Type} \El(A) \to \Type,\]
which when externalized generates the categorical description above.

The term formation rule corresponds to a morphism \(\lambda \of Q \to \Term\),
where the domain of formation is
\[Q(\Gamma) = \left\{(A, B, b)\,\middle|\,
\begin{gathered}
A \in \Type(\Gamma) \\
B \in \Type(\Gamma \cdot A) \\
b \in \Term(\Gamma \cdot A) \\
\typeof(b) = B
\end{gathered}
\right\}.\]
We require the following square to commute,
so that the terms formed by \(\lambda\) have the right type.
\[\begin{tikzcd}
Q & \Term \\
P & \Type
\arrow[from=1-1, to=2-1]
\arrow["\typeof", from=1-2, to=2-2]
\arrow["\Pi"{description}, from=2-1, to=2-2]
\arrow["\lambda"{description}, from=1-1, to=1-2]
\end{tikzcd}\]
The elimination rule and \(\beta\eta\) equality comes from
requiring such a square to be a pullback.
This completely specifies the \(\Pi\) type.

Extensional types generally have universal properties in CwFs.
Intensional types, on the other hand, are defined by \emph{lifting conditions}.
For more details, see \textcite{awodey:2017:natural}.

% Gluing in CwFs is done by
% \textcite{kaposi:2019:gluing},
% and there is a more categorical version by
% \textcite{sterling:2023:geometry}.

\subsection{Uemura's General Framework}

Each type theory determines a notion of category with families
equipped with the corresponding type structures.
This calls for a general definition.

In presheaf categories, a morphism \(\alpha \of U_* \to U\)
such that all its pullbacks with
morphisms of the form \(\yo(\Gamma) \to U\) are representable
is called a \textbf{representable natural transformation}.
So in a CwF, \(\typeof \of \Term \to \Type\) is defined to be a representable natural transformation.
The underlying category of a CwF is the semantic counterpart of contexts,
and presheaves record judgement forms.
So requiring certain natural transformations to be representable
amounts to enabling certain judgements to appear in contexts.

\textcite{uemura:2023:general} \emph{defines} a type theory
by a category with some arrows marked as ``to become representable''
(together with other relevant structures),
and a model of such a type theory is a functor into a presheaf category
that realizes the marked arrows as representable natural transformations.
This takes the perspective of Lawvere's \emph{functorial semantics}.~\cite{lawvere:1963:functorial}

For example, the type theory freely generated by a single representable arrow
produces bare CwFs, i.e. its models are CwFs without additional structure.
Uemura described how to translate a variety of type theories into this formalism,
and proved a family of important theorems such as
\emph{initiality} generalizing \Cref{lemma:initial-cwf}.

\subsection{Gratzer–Sterling Framework}

\textcite{gratzer:2021:framework} argue that
we need not track representability with type theories.
This explains the theoretical basis of our \Cref{def:model:topos}.
Instead of using the category of contexts, we use the category of judgements
as our syntactic category.
This leads to a much simpler construction of syntax,
very similar to the technique of higher-order abstract syntax,
as we shall see later.

Here, the notion of judgements is much broader than
what is usually presented.
Traditionally, there are four kinds of common judgements,
\(A\isType\), \(A = B \isType\), \(a \of A\) and \(a = b \of A\).
We are then allowed to form \emph{hypothetical} judgements
with the hypotheses restricted to judgements of the third kind,
thus producing the trademark feature of (dependent) type theories
\[a \of A, b \of B \vdash t \of C.\]
This is rather artificial, and is designed to be
the minimal structure that supports the theory.
It would be much more natural if
we require judgements to be closed under more primitive operations
such that dependent sums and products,
and build up complex judgements with them.
For example, the context would be a chain of dependent sum
of the judgements \(x \of A\),
and \(\vdash\) would be represented by a dependent product.

Not tracking representability means
not tracking what judgements count as contexts.
Thus, a type theory would be defined completely without reference to contexts.
This raises three problems.
\begin{itemize}
\item If we don't define what's a context now, when should we?
The answer is \emph{whenever we need to!}
We can simply carve out a suitable subcategory later.
This allows for the flexibility
to regard different things as contexts in different circumstances.
An example of us doing this will be seen in \Cref{subsec:dependent:normalization}.
\item How should we even define the syntax of a type theory,
if not referring to contexts?
This will be dealt with in \Cref{subsec:LF}.
\item Since we can't distinguish whether a judgement is a context,
we need to allow hypothetical judgements of arbitrarily high order such as
\[(x\Of A \vdash B \isType) \vdash C \isType.\]
Is doing this \emph{conservative}?
We need to show that it doesn't somehow lead to additional equations or exotic terms,
since HOAS does lead to these phenomena when care is not taken.
This is proved by \textcite{gratzer:2021:framework}.
\end{itemize}

\begin{idea}
We can use the category of judgements (in a generalized sense)
as our syntactic category,
to allow for more expressivity and less bureaucracy.
\end{idea}

\subsection{Category of Judgements}\label{subsec:LF}

Following \textcite{sterling:2021:thesis},
we introduce a new language, the \emph{logical framework},
for specifying type theories and generating their syntactic categories.
Note that this is different from both
the language of type theory we are trying to define,
and the internal language of the glued topoi.
To differentiate them, we call a type in the logical framework a LF-type, etc.
And we will typeset the code blocks in \colorbox{green!10}{green background}.

Types in the logical framework are called \emph{signatures}.
Some signatures are called judgements.
There is a signature \(\Jdg\) that is the LF-type of judgements.
A signature \(\mathbb S\) provides the judgement forms of a specific type theory,
and \(\mathbb S \to \Jdg\) consists of the judgements in the type theory specified by \(\mathbb S\).
Elements of \(\mathbb S\) are called \emph{implementations} of the signature,
and elements of judgements are called \emph{deductions}.
These will be clear with some examples.

We can take the dependent sum, dependent function
and extensional equality of signatures.
Judgements are closed under these operations.
There is also the trivial judgement \(\Unit\).
For notational ease, we will use record types
as a syntax sugar for nested dependent sums.

For example, here is a signature for a simple type theory.
\begin{logical}
\>[0]\kw{record} \(\mathbb S_\to\) \kw{where}\\
\>[0][@{}l@{\AgdaIndent{0}}]%
\>[4]\(\pr{Tp} \of \Jdg\)\\
\>[4]\(\pr{El} \of \pr{Tp} \to \Jdg\)\\
\>[4]\(\pr{Fun} \of \pr{Tp} \to \pr{Tp} \to \pr{Tp}\)\\
\>[4]\(\pr{app} \of \prod_{A,B \of \pr{Tp}}
    \pr{El}(\pr{Fun}(A, B)) \to \pr{El}(A) \to \pr{El}(B)\)\\
\>[4]\(\pr{lam} \of \prod_{A,B \of \pr{Tp}}
    (\pr{El}(A) \to \pr{El}(B)) \to \pr{El}(\pr{Fun}(A, B))\)
\end{logical}
We specified a judgement \(\pr{Tp}\), which is traditionally written \(A \isType\).
Similarly \(\pr{El}(A)\) corresponds to the judgement \(a \of A\).
\(\pr{Fun}\) corresponds to the function type formation rule,
which is a derivation of judgements as we said,
and similarly for \(\pr{app}\) and \(\pr{lam}\).

See how we have \(\pr{El}(A) \to \pr{El}(B)\) in the argument of \(\pr{lam}\),
which forms a kind of higher-order abstract syntax representation.
As computed earlier, this automatically leads to the correct notion of binding.

Of course we are missing \(\beta\) and \(\eta\) rules here.
We can add them in.
But for brevity we package the four constructs into one:
we write \((f, g) \of X \cong Y\) as an abbreviation for
\begin{logical}
\>[0]\(f \of X \to Y\)\\
\>[0]\(g \of Y \to X\)\\
\>[0]\(\something \of \prod_{x \of X} g(f(x)) = x\)\\
\>[0]\(\something \of \prod_{y \of Y} f(g(y)) = y\)
\end{logical}
Our complete specification of a type theory with only function types:
\begin{logical}
\>[0]\kw{record} \(\mathbb S_\to\) \kw{where}\\
\>[0][@{}l@{\AgdaIndent{0}}]%
\>[4]\(\pr{Tp} \of \Jdg\)\\
\>[4]\(\pr{El} \of \pr{Tp} \to \Jdg\)\\
\>[4]\(\pr{Fun} \of \pr{Tp} \to \pr{Tp} \to \pr{Tp}\)\\
\>[4]\((\pr{lam}, \pr{app}) \of \prod_{A,B \of \pr{Tp}}
    (\pr{El}(A) \to \pr{El}(B)) \cong \pr{El}(\pr{Fun}(A, B))\)
\end{logical}
A possible judgement in this type theory is
\[\prod_{A,B \of \pr{Tp}}\pr{El}(A) \to \pr{El}(B) \to \pr{El}(A),\]
corresponding to the more traditional
\[x \Of A, y \Of B \vdash t \of A. \quad \explain{(\(A, B\) implicitly quantified)}\]
Note that \(t\) is actually recording a derivation of the judgement, not the judgement itself.

% (...) conservative extension of \(\Sig\) universe of signatures (Do we even need them?)

\begin{definition}
Given a signature \(\mathbb S\), we define the \textbf{category of judgements}
of the type theory specified by the signature.
The objects are implementations of \(\mathbb S \to \Jdg\).
Given two objects \(J_1, J_2\), a morphism between them is
an implementation (considered up to LF-equality) of
\[\prod_{S \of \mathbb S} J_1(S) \to J_2(S).\]
Identity and composition are the obvious ones.
\end{definition}

\begin{remark}
We can equivalently define an object as an implementation of \(\Jdg\)
depending on a variable \(S \of \mathbb S\).
Similarly for the morphisms.
\end{remark}

\begin{idea}
The category of judgements of a type theory
can be defined using a logical framework.
\end{idea}

\subsection{Functorial Semantics}

The logical framework also provides us with a notion of models.
The categorical language appropriate for this is
\emph{locally Cartesian closed categories}.
In particular, topoi are LCCCs.
However, we will not actually touch the details,
because the LF-language corresponds to the internal language
in a more elementary way.
Here is a brief overview, and the reader can skip to
\Cref{sec:dependent-synth} and take its results as granted.

Roughly, a LF-context is interpreted as an object in a LCCC.
A LF-type \(\Gamma \vdash A\) is represented by the projection morphism
\((\Gamma, x \Of A) \to \Gamma\),
since they are in bijective correspondence with the types.
With extensional identity types, every substitution \(\Theta \to \Gamma\)
is isomorphic to a substitution of the form \((\Gamma, x\Of A) \to \Gamma\).
So we can also interpret a dependent LF-type as an object in the slice category over \(\Gamma\).
Substitution of types corresponds to a pullback
\[\begin{tikzcd}
{(\Theta, x \Of A[\sigma])} & {(\Gamma, x \Of A)} \\
\Theta & \Gamma
\arrow["\sigma"', from=2-1, to=2-2]
\arrow[from=1-2, to=2-2]
\arrow[from=1-1, to=1-2]
\arrow[from=1-1, to=2-1]
\arrow["\lrcorner"{anchor=center, pos=0.125}, draw=none, from=1-1, to=2-2]
\end{tikzcd}\]
which gives a functor \(\Delta_\sigma \of \mathcal C/\Gamma \to \mathcal C/\Theta\).
There is a triple adjunction
\[\begin{tikzcd}
{\mathcal C/\Gamma} && {\mathcal C/\Theta}
\arrow[""{name=0, anchor=center, inner sep=0}, "{\Delta_\sigma}"{description}, from=1-3, to=1-1]
\arrow[""{name=1, anchor=center, inner sep=0}, "{\Pi_\sigma}", curve={height=-21pt}, from=1-1, to=1-3]
\arrow[""{name=2, anchor=center, inner sep=0}, "{\Sigma_\sigma}"', curve={height=21pt}, from=1-1, to=1-3]
\arrow["\dashv"{anchor=center, rotate=90}, draw=none, from=0, to=1]
\arrow["\dashv"{anchor=center, rotate=90}, draw=none, from=2, to=0]
\end{tikzcd}\]
providing the interpretation of \(\Pi\) and \(\Sigma\) types.
The extensional equality is simply the diagonal \(A \to A \times_{\Gamma} A\).
The reader can consult \textcite{seely:1984:lccc} for a more detailed description.

Thus, to preserve the structures provided by the logical framework,
we need a locally Cartesian closed functor.
\begin{definition}
Given a signature \(\mathbb S\), a \textbf{model} of its corresponding type theory
is a LCC functor \(\mathcal T_{\mathbb S} \to \mathcal E\) from
the category of judgements to another LCCC.
A morphism of models is a LCC functor \(\mathcal E \to \mathcal F\)
that makes the triangle of functors commute.
\end{definition}

\begin{remark}
Such a definition makes the initiality property trivial.
The identity functor \(\mathcal T_{\mathbb S} \to \mathcal T_{\mathbb S}\)
represents the syntactic model,
and initiality is obvious from the constructions.
More generally, a coslice category always has an initial object.
\end{remark}

\begin{idea}
Functorial semantics provide the notion of models for
type theories generated by a signature in the logical framework.
\end{idea}

\section{Dependent Types, Synthetically}
\label{sec:dependent-synth}

We write down the signature for Martin-Löf dependent type theory.
For brevity we will omit some types that can be inferred.
\begin{logical}
\>[0]\kw{record} \(\mathbb S_{\text{ML}}\) \kw{where}\\
\>[0][@{}l@{\AgdaIndent{0}}]%
\>[4]\(\pr{Tp} \of \Jdg\)\\
\>[4]\(\pr{El} \of \pr{Tp} \to \Jdg\)\\
\\
\>[4]\(\pr{Ans} \of \pr{Tp}\)\\
\>[4]\(\pr{yes}, \pr{no} \of \pr{El}(\pr{Ans})\)\\
\\
\>[4]\(\pr{Pi} \of (A \of \pr{Tp}) (B \of \pr{El}(A) \to \pr{Tp}) \to \pr{Tp}\)\\
\>[4]\((\pr{lam}, \pr{app}) \of
    \{A, B\} \left(\prod_{a \of \pr{El}(A)} \pr{El}(B(a))\right)
    \cong \pr{El}(\pr{Pi}(A, B))\)\\
\\
\>[4]\(\pr{Sigma} \of (A \of \pr{Tp}) (B \of \pr{El}(A) \to \pr{Tp}) \to \pr{Tp}\)\\
\>[4]\((\pr{pair}, \pr{split}) \of
    \{A, B\} \left(\sum_{a \of \pr{El}(A)} \pr{El}(B(a))\right)
    \cong \pr{El}(\pr{Sigma}(A, B))\)\\
\\
\>[4]\(\pr{U} \of \pr{Tp}\)\\
\>[4]\(\pr{el} \of \pr{El}(\pr{U}) \to \pr{Tp}\)\\
\>[4]\(\pr{ans} \of \pr{El}(\pr{U})\)\\
\>[4]\(\something \of \pr{el}(\pr{ans}) = \pr{Ans}\)\\
\>[4]\(\pr{pi}, \pr{sigma} \of (A \of \pr{El}(\pr{U}))(B \of \pr{El}(\pr{el}(A)) \to \pr{El}(\pr{U})) \to \pr{El}(\pr{U})\)\\
\>[4]\(\something \of \{A, B\}\, \pr{el}(\pr{pi}(A, B)) = \pr{Pi}(\pr{el}(A), \pr{el} \comp B)\)\\
\>[4]\(\something \of \{A, B\}\, \pr{el}(\pr{sigma}(A, B)) = \pr{Sigma}(\pr{el}(A), \pr{el} \comp B)\)
\end{logical}
\begin{remark}
There is a striking resemblance of universes \(\pr{U}, \pr{el}\)
and the typing judgements \(\pr{Tp}, \pr{El}\).
This can be made more explicit,
for example by \textcite[\S{} 1.7]{sterling:2021:thesis},
where the universes are literally another pair of judgements
identical to \(\pr{Tp}, \pr{El}\),
with additional clauses stating cumulativity.
\end{remark}

\subsection{Connecting LF-world to Internal World}
We define \(\mathcal T\) to be the syntactic category generated by
\(\mathbb S_{\textrm{ML}}\).
A model of our type theory, i.e.\@ a LCC functor \(\mathcal T \to \mathcal E\),
is somewhat difficult to handle in the internal language,
similar to how proper classes are annoying in pure ZF set theory.
For example, a model in \(\C{Set}\) interprets \(\pr{Tp}\) as an arbitrary set.
Since we can't talk about the set of all sets,
this object won't have a type in the internal language.

An easier gadget to consider is a \(\iType_n\)-small model,
where everything is restricted to be under \(\iType_n\).
Now we can say \(\pr{Tp}\) is interpreted as an element of \(\iType_n\).
In this way, the signature in the LF-language can be transformed into the internal language.
\begin{lemma}
Given a universe \(\iType\) in a topos \(\mathcal E\),
a \(\iType\)-small model of \(\mathcal T\) is a record starting with
\begin{internal}
\>[0]\kw{record} Model \kw{where}\\
\>[0][@{}l@{\AgdaIndent{0}}]%
\>[4]\(\Type \of \iType\)\\
\>[4]\(\El \of \Type \to \iType\)\\
\>[4]\(\synct{Ans} \of \Type\)\\
\>[4]\dots
\end{internal}
\end{lemma}

The Yoneda embedding is a LCC functor, therefore
we automatically get a model in the topos \(\C{Psh}(\mathcal T)\),
which we can assume to be contained in some universe
using large cardinal axioms.
The details of universe fiddling requires more categorical language to make precise.
See \textcite[\S{} 3.7]{sterling:2021:thesis} for more detail.

Note that \(\yo\) does not preserve coproducts, so
the logical framework doesn't have them.
Compare this to the internal language which doesn't have this restriction,
and we can freely create inductive types such as \(\Msem X\).
% This is also why inductive types are considered intensional.

Recall that the glued topos \(\mathcal G\) contains a copy of \(\C{Psh}(\mathcal T)\)
called the \(\Msyn\)-modal types.
The model given by \(\yo\) in \(\C{Psh}(\mathcal T)\)
becomes a \(\iType_\Psyn\)-small model \(M_{\mathcal T}\) in \(\mathcal G\),
where \(\iType_\Psyn\) is the universe of \(\Msyn\)-modal \(\iType\)\!-small types.
It can be written as
\[\iType_\Psyn = \{A \of \iType \mid \text{\(A\) is \(\Msyn\)-modal}\}.\]
And as usual, we will try to construct a model
\(M_{\mathcal G}\) that restricts to \(M_{\mathcal T}\) under \(\Psyn\).

\begin{idea}
Using universes, we can compute how to
talk about models internally from the LF-signature.
\end{idea}

\subsection{Canonicity}
This follows \textcite[\S{} 4.5]{sterling:2021:thesis}.
Our glued topos \(\mathcal G = \C{Set} \CommaCat \GSect\).

We define \(\Type_{\mathcal G}\) as a G type.
\begin{internal}
\>[0]\(\Type_{\mathcal G} =
    \Refine_{A \of \Type_{\mathcal T}}
    \ext{\iType_1}{\El_{\mathcal T}(A)}\)\\
\>[0]\(\El_{\mathcal G}([A, E]) = E\)
\end{internal}
Here we use \(\iType_1\) because we need to use \(\iType_0\)
to interpret the universe inside the type theory.
A semantic type is defined to be a syntactic type equipped with
an assignment of computability structures over the syntactic terms.
\(\El_{\mathcal G}\) simply fetches the assignment.
We can verify that \(\Type_{\mathcal G}\) and \(\El_{\mathcal G}\)
restricts to \(\Type_{\mathcal T}\) and \(\El_{\mathcal T}\) under \(\Psyn\).
\begin{internal}
\>[0]\(\synct{Ans}_{\mathcal G} = [\synct{Ans}_{\mathcal T}, \hole]\)
\end{internal}
Since the first component is always forced to be the corresponding syntactic type,
we just omit it and give the second component as follows.
\begin{internal}
\>[0]\(\El_{\mathcal G}(\synct{Ans}_{\mathcal G}) =
\Refine_{t \of \El_{\mathcal T}(\synct{Ans})}
\Msem ((t = \synct{yes}) + (t = \synct{no}))\)\\
\>[0]\(\synct{yes}_{\mathcal G} = [\synct{yes}_{\mathcal T}, \ct{inr}(\ct{inl}(\ct{refl}))]\)\\
\>[0]\(\synct{no}_{\mathcal G} = [\synct{no}_{\mathcal T}, \ct{inr}(\ct{inr}(\ct{refl}))]\)
\end{internal}
where we also omit the \(\mathcal T\!/\mathcal G\) subscript if it is clear from context.
% (Maybe we should use colors.)
\begin{internal}
\>[0]\(\El_{\mathcal G}(\synct{Pi}(A, B))\)\>[4]\(=
\Refine_{f \of \El_{\mathcal T}(\synct{Pi}(A, B))}
\sum_{\tilde f \of (x \of \El_{\mathcal G}(A)) \to \El_{\mathcal G}(B(x))}\)\\
\>[4]\(\Psyn \to(\tilde f = \lambda x\bind \synct{app}_{\mathcal T}(f, x))\)
\end{internal}
These constructions are easy generalizations from the non-dependent case,
so the reader should be able to fill in the missing fields.
We now turn to the interesting part, universes.
\begin{internal}
\>[0]\(\El_{\mathcal G}(\synct{U}) =
\Refine_{A \of \El_{\mathcal T}(\synct{U})}
\ext{\iType_0}{\El_{\mathcal T}(\synct{el}(A))}\)\\
\>[0]\(\synct{el}_{\mathcal G}([A, e]) = [\synct{el}(A), e]\)\\
\>[0]\(\synct{ans}_{\mathcal G} = [\synct{ans}_{\mathcal T}, \El_{\mathcal G}(\synct{Ans})]\)
\end{internal}
Note that \(\El_{\mathcal G}(\synct{Ans})\) is indeed in the lowest universe,
and \(\synct{el}_{\mathcal G}(\synct{ans}) = \synct{Ans}_{\mathcal G}\) holds.
Let's try to define \(\synct{pi}_{\mathcal G}\).
\begin{internal}
\>[0]\(\synct{pi}_{\mathcal G}(A, B) =
    [\synct{pi}_{\mathcal T}(A^\syn, \lambda a \bind B(a)^\syn),
    \hole]\)
\end{internal}
The hole should be filled with a type in \(\iType_0\),
such that when restricted under \(\Psyn\) becomes
\(\El_{\mathcal T}(\synct{el}(\synct{pi}(A^\syn, \lambda a\bind B(a)^\syn)))\),
which is equal to
\[S = \El_{\mathcal T}(\synct{Pi}(\synct{el}(A), \lambda a\bind \synct{el}(B(a)^\syn))),\]
but only \emph{isomorphic} to \(S' = \prod_{a} \El_{\mathcal T}(\synct{el}(B(a)^\syn))\).
We fill in the hole with
\[\Refine_{f \of S} \sum_{f' \of S'} \Psyn \to
(f' = \lambda x\bind \synct{app}_{\mathcal T}(f, x)).\]
\(\synct{sigma}\) is constructed similarly.

The above constructs a model in the internal language,
which is by definition a LCC functor \(\mathcal T \to \mathcal G\) externally.
The fact that \(\Type_{\mathcal G}\) restricts to \(\Type_{\mathcal T}\) etc. under \(\Psyn\)
externally means the following diagram commutes.
\[\begin{tikzcd}
{\mathcal T} & {\mathcal G} \\
& {\C{Psh}(\mathcal T)}
\arrow[from=1-1, to=1-2]
\arrow["{\pi_\syn}", from=1-2, to=2-2]
\arrow["\yo"', from=1-1, to=2-2]
\end{tikzcd}\]
A term \(\vdash t \of \Ans\) corresponds to a morphism \(1 \to \Ans\) in the syntactic category,
which is then interpreted into \(\mathcal G\).
We finish the proof in much the same way as the non-dependent case.

\subsection{Normalization}
This follows \textcite[\S\S{} 5.5--5.6]{sterling:2021:thesis}.
\subsubsection{Carving out Contexts}\label{subsec:dependent:normalization}
We first define our topos.
By our previous experience, we know that we need to define
a subcategory \(\mathcal A \hookrightarrow \mathcal T\),
where \(\mathcal T\) is now defined using the logical framework.
We proceed inductively.
First we define what it means for an object of \(\mathcal T\)
to be a \textbf{context}.
\begin{itemize}
\item \(1\) is a context.
\item If \(\Gamma\) is a context, \(A \of \Gamma \to \Type_{\mathcal T}\) is a morphism,
then \(\sum_{x \of \Gamma} \El_{\mathcal T}(A(x))\) is a context.
\end{itemize}
Recall that objects of \(\mathcal T\) are implementations of \(\Jdg\),
depending on an assumption \(S \of \mathbb S\).
The \(\Sigma\) type here is simply the \(\Sigma\) type in the logical framework.

Next we define \textbf{variables}.
They are triples \((\Gamma, A, a)\),
where \(\Gamma\) is a context, \(A \of \Gamma \to \Type\),
and \(a \of (x \of \Gamma) \to \El(A(x))\).
We usually omit the first and second components.
\begin{itemize}
\item \(\pi_2 \of \left(x' \of \sum_{x \of \Gamma} \El(A(x))\right) \to \El(A(\pi_1(x)))\) is a variable.
\item If \((\Gamma, A, a)\) is a variable, then
\(a \comp \pi_1 \of (x' \of \sum_{x \of \Gamma} \El(B(x))) \to \El(A(\pi_1(x')))\) is a variable.
\end{itemize}
It is a mouthful, but ultimately the definition just boils down to
singling out nested projections \(\pi_2 \comp \pi_1 \comp\dots\comp\pi_1\) out of contexts.

Finally we define what it means for a morphism whose domain and codomain are contexts
to represent a \textbf{renaming}.
\begin{itemize}
\item \(\Gamma \to 1\) is a renaming.
\item If \(\sigma \of \Gamma \to \Delta\) is a renaming, and
\(a \of (x \of \Gamma) \to \El(A(\sigma(x)))\) is a variable,
then \[\lambda x \bind (\sigma(x), a(x)) \of \Gamma \to \sum_{y \of \Delta} \El(A(y))\] is a renaming.
\end{itemize}

We now have a subcategory of contexts and renamings \(\mathcal A \hookrightarrow \mathcal T\).
We now construct \(\mathcal G = \textrm{Id}_{\C{Psh}(\mathcal A)} \CommaCat \rho^*\).
This is where the synthetic language lives.

We can assemble \(\Elvar\) here.
For convenience, we will define a single object \(\Termvar\) which mixes variables of all types,
and a projection \(\typeofvar \of \Termvar \to \Type\),
just like how we have \(\typeof \of \Term \to \Type\).
We can then internally use the G type to construct \(\Elvar(A)\) for each type.

We consider a presheaf over \(\mathcal A\), defined by setting
\(V(\Gamma)\) to be the set of variables \((\Gamma, A, a)\) with the first component equal to \(\Gamma\).
The projection of the third component gives us a natural transformation
\(V \to \rho^*(\yo(\Term))\).
This is the definition of \(\Termvar\).
We have a commutative square giving \(\typeofvar\).
\[\begin{tikzcd}
V & {\rho^*(\yo(\Type))} \\
{\rho^*(\yo(\Term))} & {\rho^*(\yo(\Type))}
\arrow[from=1-1, to=2-1]
\arrow[from=2-1, to=2-2]
\arrow[from=1-2, to=2-2]
\arrow[from=1-1, to=1-2]
\end{tikzcd}\]
Now we can define
\[\Elvar(A) = \Refine_{a \of \El(A)}
\sum_{x \of \Termvar} (\typeofvar(x) = A) \land \Psyn \to (\hole = a)\]
where the hole is filled with the coercion of
the term \(x \of \Term\) to \(\El(A)\)
given that \(\typeof(x) = A\).

\begin{idea}
Contexts can be defined in the category of judgements inductively,
decoupled from the definition of the syntax itself.
\end{idea}

\subsubsection{Normal and Neutral Forms}

We also need to define normal and neutral forms.
Interestingly, we don't need to define them externally and feed them into the internal language.
They can be done entirely in the internal language itself.

We define three indexed inductive types mutually.
\(\textrm{Nf}(A, a)\) represents the \(\Msem\)-modal type of normal forms over \(a\),
where \(a \of \El_{\mathcal T}(A)\).
It is denoted as \([a \isNf A]\) for visual simplicity.
Similarly we have \([a \isNe A]\) and \([A \isNfType]\).
We don't need \([A \isNeType]\) here, because neutral types
must arise in a universe, and we can use \([A \isNe \Univ]\) instead.

To ensure the type is \(\Msem\)-modal, we simply add constructors
\(\ct{pt} \of \Psyn \to [a \isNf A]\) and
\(\ct{trunc} \of (n \of [a \isNf A])(\something \of \Psyn) \to n = \ct{pt}(\something)\).
So that under \(\Psyn\) the type is collapsed to a point.
We can then define
\[\Elnf(A) = \Refine_{a \of \El_{\mathcal T}(A)} [a \isNf A].\]
\(\Elne(A)\) and \(\Typenf\) are analogous.

\begin{internal}
\>[0]\kw{inductive} \([a \isNf A]\) \kw{where}\\
\>[0][@{}l@{\AgdaIndent{0}}]%
\>[4]\(\ct{pt} \of \Psyn \to [a \isNf A]\)\\
\>[4]\(\ct{trunc} \of (n \of [a \isNf A])(\something \of \Psyn) \to n = \ct{pt}(\something)\)\\
\>[4]\(\nfct{lam}_\sem \of (\prod_{x \of \Elvar(A)}[f(x^\syn) \isNf B(x^\syn)]) \to
    [\synct{lam}_{\mathcal T}(f) \isNf \synct{Pi}_{\mathcal T}(A, B)]\)\\
\>[4]\(\nfct{yes}_\sem \of [\synct{yes}_{\mathcal T} \isNf \synct{Ans}_{\mathcal T}]\)\\
\>[4]\dots
\end{internal}
The reader should hopefully see a pattern here.
We can now construct \(\nfct{yes} = [\synct{yes}_{\mathcal T}, \nfct{yes}_\sem]\), etc.
The other two inductive types are defined in the same way,
following \Cref{fig:dep-nf-ne}.
The constructors are also listed by \textcite[\S{} 5.4.1]{sterling:2021:thesis}.

One may worry about the quotient constructor \(\ct{trunc}\).
Since the whole point of normal forms is to make equality trivial to decide,
using quotients to define them seem questionable.
However, since the quotient only occurs under \(\Psyn\),
in the external world, the \(\Msem\) component would be an honest inductive set.

\begin{idea}
Normal and neutral forms can be defined internally using quotient-inductive types.
The quotient only truncates at \(\Psyn\) and does not affect effective computation.
\end{idea}

\subsubsection{Computability Structures}

With the addition of universes, reification and reflection are also integrated into the model,
instead of being defined afterwards.
For each syntactic type \(A \of \Type_{\mathcal T}\),
we define the type of computability structures over it.
\begin{internal}
\>[0]\kw{record} \(\Comp_i(A \of \Type_{\mathcal T})\) \kw{where}\\
\>[0][@{}l@{\AgdaIndent{0}}]%
\>[4]\(\pr{ntp} \of \ext{\Typenf}{A}\)\\
\>[4]\(\pr{sem} \of \ext{\iType_i}{\El_{\mathcal T}(A)}\)\\
\>[4]\(\reflect \of \ext{\Elne(A) \to \pr{sem}}{\lambda x\bind x}\)\\
\>[4]\(\reify \of \ext{\pr{sem} \to \Elnf(A)}{\lambda x\bind x}\)\\
\>[0]\(\Type_{\mathcal G} = \Refine_{A \of \Type_{\mathcal T}} \Comp_1(A)\)\\
\>[0]\(\El_{\mathcal G}(A) = A^\sem.\pr{sem}\)
\end{internal}
In the canonicity proof this record only consisted of the \(\pr{sem}\) component,
so we inlined it.
We will write \(\reflect^A\) for the record component
\(A^\sem.{\reflect}\), similarly for \(\reify^A\).

\begin{internal}
\>[0]\(\synct{Ans}_{\mathcal G}^\sem = \kw{record} \{\)\\
\>[0][@{}l@{\AgdaIndent{0}}]%
\>[4]\(\pr{ntp} = \nfct{Ans}\)\\
\>[4]\(\pr{sem} = \Elnf(\synct{Ans}_{\mathcal T})\)\\
\>[4]\(\reflect(a) = \nfct{up}(a)\)\\
\>[4]\(\reify(a) = a\)\\
\>[0]\(\}\)\\
\\
\>[0]\(\synct{Pi}_{\mathcal G}(A,B)^\sem = \kw{record} \{\)\\
\>[4]\(\pr{ntp} = \nfct{Pi}(A^\sem.\pr{ntp},
    \lambda a\bind B(\reflect^A \nect{var}(a))^\sem.\pr{ntp})\)\\
\>[4]\(\pr{sem} = \hole\)\\
\>[4]\(\reflect(a) = \hole\)\\
\>[4]\(\reify(a) = \hole\)\\
\>[0]\(\}\)\\
\\
\>[0]\(\synct{Sigma}_{\mathcal G} = \hole\) \explain{(Left as exercise)}
\end{internal}
These are mostly just reorganizing what we've already seen.
New stuff appears in lifting universes.
\begin{internal}
\>[0]\kw{record} \(\textrm{comp}(A \of \El_{\mathcal T}(\synct{U}))\) \kw{where}\\
\>[0][@{}l@{\AgdaIndent{0}}]%
\>[4]\(\pr{ntp} \of \ext{\Elnf(\synct{U})}{A}\)\\
\>[4]\(\pr{sem} \of \ext{\iType_0}{\El_{\mathcal T}(\synct{el}(A))}\)\\
\>[4]\(\reflect \of \ext{\Elne(\synct{el}(A)) \to \pr{sem}}{\lambda x\bind x}\)\\
\>[4]\(\reify \of \ext{\pr{sem} \to \Elnf(\synct{el}(A))}{\lambda x\bind x}\)\\
\\
\>[0]\(\synct{U}_{\mathcal G}^\sem = \kw{record} \{\)\\
\>[4]\(\pr{ntp} = \nfct{U}\)\\
\>[4]\(\pr{sem} = \Refine_{A \of \El_{\mathcal T}(\synct{U})} \mathrm{comp}(A)\)\\
\>[4]\(\reflect(A) = \hole\)\\
\>[4]\(\reify(A) = A^\sem.\pr{ntp}\)\\
\>[0]\(\}\)
\end{internal}
We need to define a function
\(\Elne(\synct{U}) \to \Refine_{A \of \El_{\mathcal T}(\synct{U})} \mathrm{comp}(A)\).
Since it is required to restrict to \(\lambda x \bind x\) under \(\Psyn\),
the first component is fixed.
The second component is
\begin{internal}
\>[0]\(\textit{helper\/}(A) = \kw{record} \{\)\\
\>[0][@{}l@{\AgdaIndent{0}}]%
\>[4]\(\pr{ntp} = \nfct{up}(A)\)\\
\>[4]\(\pr{sem} = \Elne(\synct{el}_{\mathcal T}(A^\syn))\)\\
\>[4]\(\reflect(A) = \lambda x\bind x\)\\
\>[4]\(\reify(A) = \lambda a\bind \nfct{up}(A, a)\)\\
\>[0]\(\}\)
\end{internal}
Here, the first \(\nfct{up}\) refers to the coercion of neutral type codes
to normal type codes, which is allowed since \(\Univ\) is an intensional type.
The second \(\nfct{up}\) corresponds to the last rule in \Cref{fig:dep-nf-ne}.

\subsubsection{Final Part}

We have constructed in the internal language a model,
corresponding to a LCC functor \(\mathcal T \to \mathcal G\).
This model also come equipped with reflection and reification functions.
Given a term \(\Gamma \vdash t \of A\), we again want to give the following composition.
\[\Ren{\Gamma} \longrightarrow \Sem{\Gamma} \longrightarrow
\El_{\mathcal G}(A) \longrightarrow \Elnf(A).\]
Here \(\Gamma\) is a context, and not an arbitrary object in the syntactic category.
Therefore \(\Gamma\) is also an object in \(\mathcal A\).
We have a morphism \(\yo_{\mathcal A}(\Gamma) \longrightarrow \rho^*(\yo_{\mathcal T}(\Gamma))\),
which we take as an object \(\Ren{\Gamma}\) in \(\mathcal G\).
Now we can use \(\reflect\) on the types to recursively define
\(\reflect \of \Ren{\Gamma} \to \Sem{\Gamma}\).
This completes the proof.

\subsection{Parametricity for System F}
As a bonus, we discuss parametricity for system F.
We first list the signature for the type system.
\begin{logical}
\>[0]\kw{record} \(\mathbb S_{\text{F}}\) \kw{where}\\
\>[0][@{}l@{\AgdaIndent{0}}]%
\>[4]\(\pr{Tp} \of \Jdg\)\\
\>[4]\(\pr{El} \of \pr{Tp} \to \Jdg\)\\
\>[4]\(\pr{Forall} \of (\pr{Tp} \to \pr{Tp}) \to \pr{Tp}\)\\
\>[4]\((\pr{Lam}, \pr{App}) \of \{F\}\,
(\prod_{A \of \pr{Tp}} \pr{El}(F(A))) \cong \pr{El}(\pr{Forall}(F))\)\\
\>[4]\(\pr{Fun} \of \pr{Tp} \to \pr{Tp} \to \pr{Tp}\)\\
\>[4]\((\pr{lam}, \pr{app}) \of \{A, B\}\,
(\pr{El}(A) \to \pr{El}(B)) \cong \pr{El}(\pr{Fun}(A, B))\)
\end{logical}

We construct the category
\(\mathcal G = \textrm{Id}_{\C{Psh}(\mathcal Y)} \CommaCat \GSect \times \GSect\),
whose objects are of the form \((X^L, X^R, X^\sem, X^\downarrow)\),
with
\[X^\downarrow \of X^\sem \longrightarrow \GSect(X^L) \times \GSect(X^R).\]
Note that we now have two syntactic objects at the bottom.
In our synthetic language, there are two postulated propositions
and a defined proposition.
\begin{internal}
\>[0]\kw{postulate}\\
\>[0][@{}l@{\AgdaIndent{0}}]%
\>[4]\(\Pleft \of \Omega\)\\
\>[4]\(\Pright \of \Omega\)\\
\>[4]\(\something \of \neg(\Pleft \land \Pright)\)\\
\>[0]\(\Psyn \of \Omega\)\\
\>[0]\(\Psyn = \Pleft \lor \Pright\)
\end{internal}
Externally, we interpret \(\Pleft\) as
\(0 \to \GSect(1) \times \GSect(0)\)
and \(\Pright\) as \(0 \to \GSect(0) \times \GSect(1)\).
More diagrammatically, it looks like this.
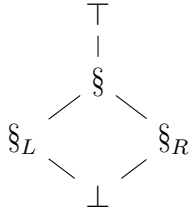
\begin{center}
\begin{tikzpicture}
\node (T) at (0,0) {\(\top\)};
\node (S) at (0, -0.9) {\(\Psyn\)};
\node (L) at (-1, -1.65) {\(\Pleft\)};
\node (R) at (1, -1.65) {\(\Pright\)};
\node (F) at (0,-2.4) {\(\bot\)};
\draw (T) -- (S);
\draw (S) -- (L) -- (F);
\draw (S) -- (R) -- (F);
\end{tikzpicture}
\end{center}
Here \(\top = 1\) is the true proposition, and \(\bot = 0\) is the false one.
Rather confusingly, \(\land\) represents the greatest lower bound,
so is depicted as a V shape in the diagram.

There are now a surfeit of modalities related to these propositions.
They have interesting relationships.
For example, \(\Msyn_L \Msyn_R X \cong 1\), and
\(\Msem_L \Msem_R X \cong \Msem X\).
Geometrically, we are looking at the three open sets
\(\{L\}, \{R\}, \{L, R\}\) of this topological space
(apart from \(\varnothing\) and the whole space).
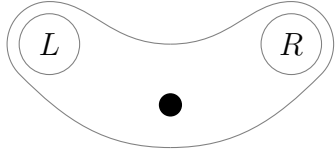
\begin{center}
\begin{tikzpicture}[scale=0.8]
\node at (-2, 0) {\(L\)};
\node at (2, 0) {\(R\)};
\node at (0, -1) {\(\Msem\)};
\draw[gray] (-2,0) circle (0.5);
\draw[gray] (2,0) circle (0.5);
\draw[gray] (0,0) .. controls (1,0) and (1.3,0.7) .. (2,0.7)
    arc (90:-45:0.7) to[in=0, out=-135] (0,-1.7);
\draw[gray] (0,0) .. controls (-1,0) and (-1.3,0.7) .. (-2,0.7)
    arc (90:225:0.7) to[in=180, out=-45] (0,-1.7);
\end{tikzpicture}
\end{center}
We try to assign some object to each of the points.%
\footnote{Technically these are not points because they may have additional inner structure.
If we glue \(\C{Set}\) onto \(\C{Set} \times \C{Set}\),
then they will truly be points.
For more geometric connections, see the entire Part C of \textcite{johnstone:2008:elephant}.}
Since the bottom point is not separable from \(L\) and \(R\),
the assignment is not independent, which explains the function \(X^\downarrow\).
\(\Msyn_L\) erases all information outside of the open set \(\{L\}\),
and \(\Msem_L\) erases all information outside of the closed set \(\{\Msem, R\}\).

\begin{center}
\begin{tabular}{c|c!{\(\longrightarrow\)}r@{\({}\times{}\)}l}
\(X\) & \(X^\sem\) & \(\GSect(X^L)\) & \(\GSect(X^R)\)\\
\(\Msyn_L X\) & \(\GSect(X^L)\) & \(\GSect(X^L)\) & \(\GSect(1)\)\\
\(\Msem_L X\) & \(X^\sem\) & \(\GSect(1)\) & \(\GSect(X^R)\)\\
\(\Msyn X\) & \(\GSect(X^L) \times \GSect(X^R)\) & \(\GSect(X^L)\) & \(\GSect(X^R)\)\\
\(\Msem X\) & \(X^\sem\) & \(\GSect(1)\) & \(\GSect(1)\)
\end{tabular}
\end{center}

Since propositions are proof-irrelevant, when we case-split on a disjunction,
we have to prove the results agree when both branch holds.
But since \(\Pleft \land \Pright\) is false, we are spared the trouble.
We will write \(\Casing{x}{y}\) for this.
We also revise our extension types to
\[\extt{A}{a_L}{a_R},\]
which can be defined in terms of the old extension type and 
a case-splitting
\(\ext[{\Pleft \lor \Pright}]{A}{\Casing{a_L}{a_R}}\).

We also need an updated version of gluing.
We form the type
\[\Refine_{x \of A, y \of B} C(x, y) \quad \text{or} \quad
\Refine_{x \of A}^{y \of B}C(x,y)\]
where \(A\) is \(\Msyn_L\)-modal, \(B\) is \(\Msyn_R\)-modal
and \(C\) is \(\Msem\)-modal.
This type is isomorphic to \(\sum_{x \of A} \sum_{y \of B} C(x, y)\),
but when \(\Pleft\) holds it is equal to \(A\),
and when \(\Pright\) holds it is equal to \(B\).
We write \(p = [x, y, z]\) for an element of this type,
and \(p^L, p^R, p^\sem\) for the three projections.
But note we use the subscripts \(L,R\) as only part of the names.

The syntax provides us with \emph{two} copies of models,
one is in the \(\Msyn_L\)-modal universe, and the other in the \(\Msyn_R\)-modal one.
We write \(\Type_L, \Type_R\) etc. to differentiate.
Our goal is to use these to construct a model that restricts
to the left model when \(\Pleft\) holds, and the right model when \(\Pright\) holds.

System F is very annoying because of its impredicativeness.
We cannot use normal approach, defining 
\[\Type_{\mathcal G} = \Refine_{A_L \of \Type_L}^{A_R \of \Type_R}
    \extt{\iType_0}{\El_L(A_L)}{\El_R(A_R)}\]
which bumps the universe level.
Instead, we use the impredicativity of propositions.
Denote \(\Omega_{\setminus \Psyn}\) the type of \(\Msem\)-modal propositions.
\begin{internal}
\>[0]\(\Type_{\mathcal G} = \Refine_{A_L \of \Type_L}^{A_R \of \Type_R}
    \El_L(A_L) \to \El_L(A_R) \to \Omega_{\setminus \Psyn}\)\\
\>[0]\(\El_{\mathcal G}([A_L, A_R, E]) =
\Refine_{a_L \of \El_L(A_L)}^{a_R \of \El_R(A_R)} E(a_L, a_R)\)\\
\>[0]\(\synct{Fun}_{\mathcal G}(A, B)= [\synct{Fun}_L(A^L, B^L), \synct{Fun}_R(A^R, B^R), \hole]\)
\end{internal}
Again, since the first two components are always fixed, we omit them and directly write
\begin{internal}
\>[0]\(\synct{Fun}_{\mathcal G}(A, B)^\sem \)
\>[4]\(=\lambda f_L, f_R\bind (a \of \El_{\mathcal G}(A)) \to\)\\
\>[4]\(B^\sem(\synct{app}_L(f_L, a^L), \synct{app}_R(f_R, a^R))\)\\
\>[0]\(\synct{lam}_{\mathcal G}(f)^\sem = \lambda a \bind
f(a)^\sem\)\\
\>[0]\(\synct{app}_{\mathcal G}(f, a)^\sem = f^\sem(a)\)
\end{internal}
We can deal with polymorphism similarly.
\begin{internal}
\>[0]\(\synct{Forall}_{\mathcal G}(F)^\sem\)\>[4]\(= \lambda t_L, t_R\bind
(A \of \Type_{\mathcal G}) \to\)\\
\>[4]\(F(A)^\sem(\synct{App}_L(t_L, A^L), \synct{App}_R(t_R, A^R))\)\\
\>[0]\(\synct{Lam}_{\mathcal G}(t)^\sem = \lambda A \bind t(A)^\sem\)\\
\>[0]\(\synct{App}_{\mathcal G}(t, A)^\sem = t^\sem(A)\)
\end{internal}

This defines a model, which gives an LCC functor \(\Sem{-} \of \mathcal T \to \mathcal G\).
Now we have \emph{two} projection functors \(\pi_{L/R} \of \mathcal G \to \C{Psh}(\mathcal T)\),
so by initiality we know that \(\pi_{L/R} \circ \Sem{-} = \yo\), and
the two syntactic parts will always be equal.
This is the main result of parametricity.

Let's try the ``hello world'' example of parametricity:
consider the type \(\forall X\bind X \to X\).
For any closed term \(t\) with this type,
we have an interpretation
\(\Sem{t} \of \El_{\mathcal G}(\synct{Forall}_{\mathcal G}(\lambda X\bind \synct{Fun}_{\mathcal G}(X, X)))\).
Let's expand the definitions. We use \(L/R\) to avoid code duplication.
\begingroup
\allowdisplaybreaks
\begin{align*}
&\phantom{{}={}}\El_{\mathcal G}(\synct{Forall}_{\mathcal G}(\lambda X\bind \synct{Fun}_{\mathcal G}(X, X)))\\
&= \Refine_{t_L}^{t_R} \synct{Forall}_{\mathcal G}(\lambda X\bind \synct{Fun}_{\mathcal G}(X, X))^\sem (t_L, t_R) \\
&= \Refine_{t_L}^{t_R} \prod_{A \of \Type_{\mathcal G}}
    \synct{Fun}_{\mathcal G}(A, A)^\sem(\synct{App}_{L/R}(t_{L/R}, A^{L/R}))\\
&= \Refine_{t_L}^{t_R} \prod_{A \of \Type_{\mathcal G}} \prod_{a \of \El_{\mathcal G}(A)}
A^\sem(\synct{app}_{L/R}(\synct{App}_{L/R}(t_{L/R}, A^{L/R}), a^{L/R}))
\end{align*}
\endgroup
By the definition of \(\Type_{\mathcal G}\) we see that
\(A\) is two closed syntactic types \(A^L, A^R\) together with a relation
\(A^\sem\) that relates the closed terms of \(A^L\) and \(A^R\).
\(a \of \El_{\mathcal G}(A)\) is simply a pair \((a^L, a^R)\) that happen to be related by \(A^\sem\).
(Recall that terms are considered up to \(\beta\eta\) equality.)
We know that \(\Sem{t}^L = \Sem{t}^R = t\).
The inner component is basically saying that the relation \(A^\sem\)
relates the terms \(t A^L a^L\) and \(t A^R a^R\).

\section{Prospect}

Although lengthy, these notes only covered proofs for well-known results.
Nonetheless synthetic Tait computability also directly contributed to novel theorems.
Normalization for cubical type theories is proved using this potent framework.~\cite{sterling:2021:cubical}
On the practical side,
a mechanism to control definition unfolding in type theory
is closely related to, and has metatheorems established by
synthetic Tait computability.~\cite{gratzer:2022:controlling}
STC can also be fruitfully applied to the study of \emph{phase distinctions} in computer science.
It isn't a great stretch to call STC a language for phase distinctions.

There are various technical aspects that I did not cover,
for example how to add an elimination principle to \(\Ans\) making it an inductive type.
Also, negative connectives can be glued in a uniform way,
but we constructed them manually every time.
The reader can consult \textcite{sterling:2021:thesis}.

One may expect to extract an algorithm from the normalization proof.
But due to the use of G types, this is not immediately clear,
since \textcite{gratzer:2022:strict} used classical principles to produce its interpretation.
Of course, there is a slick way to recover a very inefficient algorithm:
since normal forms exist, one can simply enumerate all valid deductions in the type system
until reaching a proof of $t = t_0$ with $t_0$ normal.
Additionally, it is possible to give constructive interpretation
for a restricted usage of the G type~\cite{gratzer:2022:controlling},
and in our simple demonstration we did not protrude from such confines.

The mechanisms of synthetic Tait computability
can be approximated by existing structures in Agda,
but it is far from perfect.
For instance,
we can use cofibrations and glue types in cubical Agda
to simulate propositions and realignment types in STC.
We have the judgemental equalities but not
the correct notion of disjunctions of propositions.
We can alternatively use the proof irrelevant \(\textsf{Prop}\) type,
postulating the equalities as axioms,
but they are tedious to use.
It would be useful to implement proof assistants that follow
the precise rules in STC.

\emergencystretch=1em
\printbibliography

\end{document}